\documentclass[11pt,a4paper]{article}

\usepackage[colorlinks=true, linkcolor=black!50!blue, urlcolor=blue, citecolor=blue, anchorcolor=blue]{hyperref}
\usepackage[font=small,labelfont=bf,margin=0mm,labelsep=period,tableposition=top]{caption}
\usepackage[a4paper,top=3cm,bottom=2.5cm,left=2.5cm,right=2.5cm,bindingoffset=0mm]{geometry}

\usepackage{graphicx}
\usepackage{float}
\usepackage{afterpage}
\usepackage{epsfig,cite}
\usepackage{amssymb}
\usepackage{amsmath}
\usepackage{multirow}
\usepackage{url}
\usepackage{xcolor}
\usepackage{float}
\usepackage{afterpage}

\usepackage{url}

\usepackage{booktabs}
\usepackage{tikz}
\usetikzlibrary{arrows}
\usepackage{tikz-3dplot}
\usepackage{enumitem}
\usepackage{hyperref}
\usepackage{cite}
\graphicspath{{./figs/}}

\def\smallfrac#1#2{\hbox{$\frac{#1}{#2}$}}
\newcommand{\lp}{\left(}
\newcommand{\rp}{\right)}
\newcommand{\bare}{{(0)}}
\newcommand{\sym}{\mathrm{sym}}
\newcommand{\asy}{\mathrm{asy}}
\newcommand{\as}{\alpha_s}

\newcommand{\nsv}{\mathrm{V}_3}
\newcommand{\nst}{\mathrm{T}_3}
\newcommand{\eg}{{\em e.g.}}
\newcommand{\ie}{{\em i.e.}}

\numberwithin{equation}{section}
\numberwithin{figure}{section}
\numberwithin{table}{section}

\usepackage{tabularx}
\newcolumntype{C}[1]{>{\centering\arraybackslash}p{#1}}

\begin{document}
\newgeometry{top=1.5cm,bottom=1.5cm,left=2.5cm,right=2.5cm,bindingoffset=0mm}

\vspace*{.2cm}

\begin{center}
{\Large \bf Parton distributions from lattice data:\\[0.2cm]
{\Large the nonsinglet case}}
\vspace{1.4cm}

 {\small
Krzysztof Cichy$^{1}$, 
Luigi Del Debbio$^{2}$ and
Tommaso Giani$^{2}$
}

\vspace{1.0cm}
 
{\it \small 
         ~$^1$ Faculty of Physics, Adam Mickiewicz University,\\
  Umultowska 85, 61-614 Pozna´n, Poland \\
         ~$^2$ The Higgs Centre for Theoretical Physics, The University of Edinburgh,\\
  Peter Guthrie Tait Road, Edinburgh EH9 3FD, United Kingdom\\[0.1cm]
}

\vspace{1.0cm}
{\bf \large Abstract}
\end{center}

{\noindent We revise the relation between Parton Distribution Functions (PDFs)
and matrix elements computable from lattice QCD, focusing on the quasi-Parton
Distribution Functions (qPDFs) approach.
We exploit the relation between PDFs and qPDFs in the case of the unpolarized
isovector parton distribution to obtain a factorization formula relating the
real and imaginary part of qPDFs matrix elements to specific nonsinglet
distributions, and we propose a general framework to extract PDFs from the
available lattice data, treating them on the same footing as experimental data.
We implement the proposed approach within the {\tt NNPDF} framework, 
and we study the potentiality of such lattice data in constraining PDFs, assuming some plausible scenarios to assess the unknown systematic uncertainties.
We finally extract the two nonsinglet distributions involved in our analysis from a selection of the available lattice data.
}

\clearpage
\tableofcontents

\section{Introduction}
\label{sec:introduction}
A precise understanding of the structure of the proton, encoded in the Parton
Distribution Functions (PDFs), is required to make predictions and analyses in
collider physics. The PDFs sets currently used in phenomenology studies are
extracted from global QCD analysis over experimental data \cite{Ball:2017nwa,
Dulat:2015mca, Alekhin:2017kpj, Martin:2009iq, Buckley:2014ana}. The increasing
number and precision of the experiments included in these fits, together with
the development of robust fitting methodologies, are pushing the determination
of PDFs to a new level of accuracy.

PDFs are formally defined as matrix elements of renormalized operators in QCD.
The computations of PDFs from first principle using lattice QCD has been
considered a great challenge for a long time. However, recent theoretical
developments~\cite{PhysRevLett.110.262002,Ji:2014gla} have shown how lattice QCD
simulations of certain spatial correlation functions between two boosted nucleon
states can give access to the so-called quasi-PDFs. These functions are then
directly related to PDFs by means of a factorization theorem, just like
high-energy processes cross sections are.

Following these novel ideas, a number of publications have appeared (see
Refs.~\cite{Cichy:2018mum,Monahan:2018euv} for recent reviews), from both
lattice and high-energy communities, addressing the main theoretical problems of
the new lattice approach: the definition and renormalization of the non-local
operators involved in the lattice simulation
\cite{Ji:2015jwa,Ishikawa:2016znu,Constantinou:2017sej,Alexandrou:2017huk,Ji:2017rah,Ji:2017oey,Ishikawa:2017faj,Green:2017xeu,Radyushkin:2018nbf,Zhang:2018diq,Li:2018tpe},
the proof of the factorization theorem between PDFs and quasi-PDFs
\cite{Xiong:2013bka,Ji:2014gla,Ma:2014jla,Briceno:2017cpo,Ma:2017pxb,Izubuchi:2018srq},
the computation of the matching coefficients relating lattice-computable
quantities to PDFs in different renormalization schemes
\cite{Xiong:2013bka,Ma:2014jla,Ji:2015qla,Xiong:2015nua,Ma:2017pxb,Wang:2017qyg,Stewart:2017tvs,Izubuchi:2018srq,Alexandrou:2018pbm,Alexandrou:2018eet,Liu:2018uuj,Liu:2018hxv}.
Also, data coming from first lattice QCD simulations have started appearing and
have gotten into already a relatively advanced stage over the last few years
\cite{Lin:2014zya,Alexandrou:2015rja,Chen:2016utp,Alexandrou:2016jqi,Zhang:2017bzy,Alexandrou:2017huk,Lin:2017ani,Chen:2017gck,Alexandrou:2018pbm,Chen:2018xof,Chen:2018fwa,Alexandrou:2018eet,Liu:2018uuj,Lin:2018qky,Fan:2018dxu,Liu:2018hxv,Alexandrou:2019lfo,Izubuchi:2019lyk}.
This gives an idea of what PDFs from the lattice might look like, not only for
nonsinglet quark PDFs of the nucleon, but also for the pion PDF and distribution
amplitude, as well as for the gluon PDF of the nucleon. Given the general
interest shown by the community, a quick improvement in the technologies
involved in such lattice simulations is to be expected in the next few years. A
great quantity of increasingly precise lattice data is then likely to be
available in the near future, requiring detailed studies about the possible
impact they might have on the overall precision of PDFs determination.

Despite the increasing number of numerical results becoming available, an
optimal strategy for reconstructing the PDFs from these data has not been
entirely addressed yet. For example, in Ref.~\cite{Karpie2019} a series of
possible approaches to tackle the problem of incomplete and discretized Fourier
transform has been presented, showing some interesting and promising results. In
this work, we exploit the factorization of quasi-PDFs in PDFs and perturbatively
computable coefficients to extract nonsinglet distributions from the data of
Refs.~\cite{Alexandrou:2018pbm,Alexandrou:2019lfo}, setting a general strategy
within the {\tt NNPDF} framework, which allows to systematically extract parton
distributions from the available lattice data. The discussion is focused on the
quasi-PDFs case, but it can be extended to include data coming from different
lattice approaches, like for example pseudo-PDFs
\cite{Radyushkin:2017cyf,Orginos:2017kos,Karpie:2018zaz}, fictitious heavy/light
quark~\cite{Detmold:2005gg,Braun:2007wv} or current-current correlators
\cite{Ma:2014jla,Ma:2017pxb,Sufian:2019bol}, paving the way to a first global
lattice QCD analysis~\cite{Ma:2017pxb}.

The structure of the paper is the following. In Sec.~\ref{sec:theoryoverview} we
revise the theoretical basis of the problem and we set up the notation. In
Sec.~\ref{sec:PDFstoqPDFs} we describe the lattice data considered in this work,
together with the the corresponding systematic uncertainties. We consider
different possible scenarios for the latter, studying their impact on the final
result. We then work out a factorization formula relating such data to the two
nonsiglet distributions $T_3$ and $V_3$, defined in
Sec.~\ref{sec:theoryoverview}. In Sec.~\ref{sec:fit} we describe the fitting
settings, revising the main feature of the {\tt NNPDF} framework and finally in
Sec.~\ref{sec:results} we present our results. We conclude in
Sec.~\ref{sec:summary}.

\newcommand{\MSb}{\overline{\mathrm{MS}}}
\newcommand{\MMSb}{{\rm M}\overline{\mathrm{MS}}}
\section{From Parton Distributions to quasi-Parton Distributions}
\label{sec:theoryoverview}
In this section, we revise the formal definition of PDFs and quasi-PDFs as
matrix elements of non-local operators in QCD. We recall the main theoretical
framework concerning the definition, renormalization and matching between
spatial and light-cone quantities. For a broad overview of several aspects of
quasi-PDFs, we refer to the recent extensive review~\cite{Cichy:2018mum}, and
references therein.

The bare unpolarized quark PDF is defined
as~\cite{Collins:1980ui,Collins:1981uw} 
\begin{align}
	\label{eq::barepdf}                                                  
	f_\mathrm{q}^\bare\lp x \rp = \lp 2P^+\rp\, \int \frac{d\xi^-}{4\pi} e^{-ixP^+\xi^-} 
	\langle \text{P}|\bar{\psi}_\mathrm{q}^\bare\lp\xi^-\rp\gamma^+ \,   
	U\lp\xi^-,0\rp \psi_\mathrm{q}^\bare\lp 0\rp  |\text{P}\rangle\, ,   
\end{align}
where $|\text{P}\rangle$ denotes a hadronic state with momentum $P^\mu = \lp
P^0,0,0,P^z\rp$, and $P^{\pm}=\smallfrac {\lp P^0 \pm P^z \rp}{ \sqrt{2}}$ are
light-cone coordinates. The index $\mathrm{q}$ identifies the parton under
investigation. For instance, in a theory where we only consider the four
lightest quarks, we have $q=u,d,s,c$. The momentum carried by the parton is
$k^\mu = x P^\mu$, $\psi_\mathrm{q}^\bare$ is the bare quark field operator and the
Wilson line $U$ is given by 
\begin{align}
	\label{eq::wilsonline}                                                      U\lp\xi^-,0\rp = \text{P}\exp 
    \lp -ig\int_0^{\xi^-}d\eta^- A^{\bare\,+}\lp \eta^- \rp \rp\, .         
\end{align}
An analogous definition can be given for the gluon bare PDFs, denoted as
$f_g^\bare\lp x \rp$. The superscripts $\bare$ in the above expressions identify
bare fields: the matrix elements that enter in the definition of
$f_\mathrm{q}^\bare$ are ultraviolet divergent, and therefore need to be
renormalized. Renormalized parton distributions are usually defined by minimal
subtraction, and the relation between the bare and the renormalized quantities
is given by
\begin{align}
	\label{eq:RenormPDF}                                   
	f_a^\bare\lp x \rp = \sum_{b}\int_x^1\frac{dy}{y}\,\text{Z}_{ab}\lp\frac{x}{y},\mu \rp f_b\lp y,\mu^2 \rp\, , 
\end{align}
where the indices $a$ and $b$ run over all the parton types (gluon and flavors
of quarks) and $\mu$ denotes the renormalization scale introduced by the minimal
subtraction scheme~\cite{Collins:1980ui}. 

Focusing on the quark PDFs for now, the renormalized distributions introduced
above have a compact support given by the interval $[-1,1]$. For
phenomenological applications, it is customary to define the PDFs on the
interval $[0,1]$, and to introduce independent functions for the quarks and the
antiquarks, which we denote $q(x,\mu^2)$ and $\bar{q}(x,\mu^2)$ respectively.
The relation between $f_q$, $q$ and $\bar{q}$ is 
\begin{equation}
    \label{eq:DefFQQbar}
    f_\mathrm{q}\lp x,\mu^2\rp = 
    \begin{cases}
        \phantom{-}q(x,\mu^2)\, , &\quad \mathrm{if}\ x>0\, , \\
        -\bar{q}(-x,\mu^2)\, , &\quad \mathrm{if}\ x<0 \, .
    \end{cases}
\end{equation}
It is useful to consider the symmetrised and antisymmetrised combinations of
$f_\mathrm{q}$ in the interval $x\in[0,1]$:
\begin{eqnarray}
	\label{eq:fsym}
	f^\mathrm{sym}_\mathrm{q}(x,\mu^2)  &= f_\mathrm{q}(x,\mu^2) + f_\mathrm{q}(-x,\mu^2) 
	\, , \\
	\label{eq:fasym}
	f^\mathrm{asy}_\mathrm{q}(x,\mu^2)  &= f_\mathrm{q}(x,\mu^2) - f_\mathrm{q}(-x,\mu^2) \, .
\end{eqnarray}
It can be readily shown that
\begin{align}
    f^\sym_\mathrm{q}(x,\mu^2) &= 
    q(x,\mu^2) - \bar{q}(x,\mu^2) = q^-(x,\mu^2) \, , \\
    f^\asy_\mathrm{q}(x,\mu^2) &= 
    q(x,\mu^2) + \bar{q}(x,\mu^2) = q^+(x,\mu^2) \, .
\end{align}
where $q^+$ and $q^-$ are defined by the equations above. The flavor
decomposition can be rewritten by collecting the quark fields in a vector, \eg\
$\psi = \lp \psi_u,\psi_d,\psi_s,\psi_c\rp$, and defining the following nonsinglet bare
PDFs:
\begin{eqnarray}
    \label{eq:fADef}
    f_A^\bare(x) &= \int \frac{d\xi^-}{4\pi} e^{-ixP^+\xi^-} 
	\langle \text{P}|\bar{\psi}^\bare\lp\xi^-\rp \lambda_A \gamma^+ \,   
	U\lp\xi^-,0\rp \psi^\bare\lp 0\rp  |\text{P}\rangle\, , 
\end{eqnarray}
where $A=3,8,15$, and we have used the Gell-Mann matrices
\begin{eqnarray}
    \lambda_3=
    \begin{pmatrix}
        1 & 0 & 0 & 0\\
        0 & -1& 0 & 0\\
        0 & 0 & 0 & 0\\
        0 & 0 & 0 & 0
    \end{pmatrix}\, , \quad
    \lambda_8=
    \begin{pmatrix}
        1 & 0 & 0 & 0\\
        0 & 1& 0 & 0\\
        0 & 0 & -2 & 0\\
        0 & 0 & 0 & 0
    \end{pmatrix}\, , \quad
    \lambda_{15}=
    \begin{pmatrix}
        1 & 0 & 0 & 0\\
        0 & 1& 0 & 0\\
        0 & 0 & 1 & 0\\
        0 & 0 & 0 & -3
    \end{pmatrix}\, . 
\end{eqnarray}
In this notation $f_3$ corresponds to $f^{u-d}=f_u-f_d$, $f_8=f^{u+d-2s}$, and
so on. The symmetrised and antisymmetrised combinations map directly into the
so-called {\em evolution basis} for the PDFs that is normally used in
phenomenological studies, see \eg\ Ref.~\cite{Vogt:2004ns} for a detailed
definition of the flavor decomposition. More specifically, we have:
\begin{align}
    f^\asy_{3}  &= u^+ - d^+ = T_3 \, , \\
    f^\sym_{3}  &= u^- - d^- = V_3 \, , \\
    f^\asy_{8}  &= u^+ + d^+ - 2 s^+ = T_8 \, , \\
    f^\sym_{8}  &= u^- + d^- - 2 s^- = V_8 \, , \\
    f^\asy_{15} &= u^+ + d^+ + s^+ - 3 c^+ = T_{15} \, , \\
    f^\sym_{15} &= u^- + d^- + s^- - 3 c^- = V_{15} \, .
\end{align}
We will return to these relations below when we discuss the factorization of the
quasi-PDFs. The renormalized PDFs $f_{\text{a}}\lp x,\mu^2\rp$ -- or
equivalently $q(x,\mu^2)$ and $\bar{q}(x,\mu^2)$ -- are currently
extracted from global fits of experimental data, see \eg\
Refs.~\cite{Gao:2017yyd,Lin:2017snn} for recent reviews.

Eq.~\eqref{eq::barepdf} defines the PDFs in terms of matrix elements of QCD
operators computed between hadron states, which makes their universal and
non-perturbative nature explicit. Their direct computation through lattice QCD
simulations looks therefore a very appealing target. Unfortunately, the real time
dependence of PDFs makes it impossible to compute these matrix elements on the
lattice, where the theory is defined on a Euclidean space with imaginary time.
To overcome this problem, a new approach was proposed in Ref.~\cite{PhysRevLett.110.262002}:
the matrix element appearing in Eq.~\eqref{eq::barepdf}, defined along a
light-cone direction, is replaced by a correlator defined along a purely spatial
direction. The resulting quantity is called a (bare) quasi-PDF.
Denoting by $\Gamma$ a generic Dirac structure and by the suffix $A$ the specific nonsinglet distribution
we want to consider, we can introduce the {\em Ioffe time distributions} \cite{Radyushkin:2017cyf,Braun:1994jq}, 
defined as the matrix element between nucleon states with momentum $P$
\begin{align}
	\label{eq:Ioffe}
	M^\bare_{\Gamma,A}\left(\xi,P\right) &= \langle P |\mathcal{M}^\bare_{\Gamma,A}\left(\xi\right) |P\rangle \, ,
\end{align}
with
\begin{align}
	\label{eq:bilocal}
	\mathcal{M}^\bare_{\Gamma,A}\left(\xi\right)= \bar{\psi}^\bare\lp\xi\rp \lambda_A \Gamma \,   
	U\lp\xi,0\rp \psi^\bare\lp 0\rp.
\end{align}
The vector bilocal operator obtained for $\Gamma=\gamma^\mu$ can be decomposed
in terms of two form factors depending on Lorentz invariants:
\[
    \label{eq:IoffeDecomposition}
    M^\bare_{\gamma^\mu,A}\left(\xi, P\right)
    = 2 P^\mu\,\text{h}^{(0)}_{\gamma^\mu,A}\lp \xi\cdot P,\xi^2 \rp
    + \xi^\mu\, \tilde{\text{h}}^{(0)}_{\gamma^{\mu},A}\lp \xi\cdot P,\xi^2 \rp \, . 
\]
The light-cone PDF in Eq.~\eqref{eq:fADef} is obtained by taking the Fourier
transform with respect to $\xi^-$ of a Ioffe time distribution with $\Gamma =
\gamma^+$, $\xi=\left(0,\xi^-,0_{\perp}\right)$ and $P=(P^+,0,0_{\perp})$, given
by
\begin{align}
	\label{eq:Ioffelightcone}
    M^\bare_{\gamma^+,A}\left(\xi,P\right) =  
    \langle \text{P}|\bar{\psi}^\bare\lp\xi^-\rp \lambda_A \gamma^+ \,   
	U\lp\xi^-,0\rp \psi^\bare\lp 0\rp  |\text{P}\rangle
	= 2 P^+\,\text{h}^{(0)}_{\gamma^+,A}\lp \xi^- P^+,0 \rp\, .
\end{align} 
Choosing instead a pure spatial direction $\xi=\left(0,0,0,z\right)$, and taking
the Fourier transform with respect to $z$, we obtain the definition of the
quasi-PDF  
\begin{align}
	\label{eq::bareqpdf}                                                 
	\tilde{f}_A^{(0)}\lp x, P_z\rp = 
	\int_{-\infty}^{\infty}\frac{dz}{4\pi}\,e^{i x \text{P}_z z}\,M^\bare_{\Gamma,A}\left(z,P\right)\, .
\end{align}
Choosing for example $\Gamma = \gamma^0$ we get
\begin{align}
	\label{eq::bareqpdf}                                                 
	\tilde{f}_A^{(0)}\lp x, P_z\rp = 
	2 P_0 \, \int_{-\infty}^{\infty}\frac{dz}{4\pi}\,e^{i x \text{P}_z z}\,\text{h}^{(0)}_{\Gamma,A}\lp z P_z,z^2 \rp, 
\end{align}
which will be the case considered in this work. As in the case of standard PDFs
(which from now on we will call {\em light-cone}\ PDFs), the matrix elements
defining the quasi-PDFs contain UV divergences, and need to be renormalized. The
perturbative renormalization of a bilocal operator, as the one appearing in
Eq.~\eqref{eq::bareqpdf}, is one of the theoretical problems of this novel
approach. It was shown in Ref.~\cite{Ishikawa:2017faj}, that the UV behaviour of
the quasi-PDFs is very different from that of PDFs, and that the position space
operator appearing in Eq.~\eqref{eq::bareqpdf} can be multiplicatively
renormalized, according to
\begin{align}
    \text{h}_{\Gamma,A}\lp z P_z,z^2,\mu \rp = Z_{A}(z)\,
    e^{\delta m |z|} \text{h}^\bare_{\Gamma,A}\lp z P_z,z^2 \rp, 
\end{align} 
where the exponential factor $e^{\delta m |z|}$ reabsorbs the power divergence
from the Wilson line, and the position-dependent factor $Z_{A}(z)$ takes care of
the remaining UV logarithmic divergences.
Importantly, the quasi-PDFs retain a dynamical dependence on the hadron momentum
$P$, unlike PDFs, which are defined to be invariant under Lorentz boosts. Also,
their support is defined to be the full real axis.

The interest in quasi-PDFs comes from the potential to relate them to light-cone
PDFs; factorization allows us to rewrite the quasi-PDFs as a convolution of the
light-cone PDFs with a coefficient function that can be computed in perturbation
theory, up to corrections that are suppressed by inverse powers of $P_z$. As
verified at 1-loop order in Ref.~\cite{Xiong:2013bka}, and proved in
Refs.~\cite{Ma:2017pxb,Izubuchi:2018srq} using the OPE, renormalized quasi-PDFs share the
same IR behaviour as the renormalized light-cone PDFs. It follows that they can
be written as
\begin{align}
	\label{eq::pdftoqpdf}                                                                             
	\tilde{f}_A\lp x , {\mu}^2 \rp =                                                               
	\int_{-1}^{1} \frac{dy}{|y|}\, C_A\lp\frac{x}{y},\frac{\mu}{P_z},\frac{\mu}{\mu'} \rp  f_A\lp y, {\mu'}^2\rp 
	+ \mathcal{O}\lp \frac{M^2}{P_z^2},\frac{\Lambda^2_{\text{QCD}}}{P_z^2} \rp\, ,                   
\end{align}
where the terms $\mathcal{O}\lp
\smallfrac{M^2}{P_z^2},\smallfrac{\Lambda^2_{\text{QCD}}}{P_z^2} \rp $ include
the power corrections suppressed by the hadron momentum. The functions $C_A$,
usually called matching coefficients, depend on the choice of the
renormalization scheme, and on the kind of quasi-PDF under consideration. The
first matching expressions, for all Dirac structures, were derived in
Ref.~\cite{Xiong:2013bka}, using a simple transverse momentum cutoff scheme. In
later works, matching coefficients were derived that relate the quasi-PDFs in
different renormalization schemes to light-cone PDFs in the $\MSb$ scheme. The
matching from $\MSb$ quasi-PDFs was first considered in
Ref.~\cite{Wang:2017qyg}, both for non-singlet and singlet quark PDFs, as well
as for gluons. Even though one can choose operators for the latter that do not
mix with singlet quark quasi-PDFs under
renormalization~\cite{Zhang:2018diq,Li:2018tpe}, mixing under matching is
inevitable. 
No mixing of the flavour nonsinglet sector with flavour singlet or
gluon sectors occurs, as stated in Eq.~\eqref{eq::pdftoqpdf}. Ref.~\cite{Wang:2017qyg} did not, however, address the
known issue of self-energy corrections, exhibiting a logarithmic UV divergence.
This was resolved in Ref.~\cite{Izubuchi:2018srq} by adding terms outside of the
plus prescription in the matching coefficient. As noticed in
Ref.~\cite{Alexandrou:2018pbm}, such prescription for renormalizing this
divergence violates vector current conservation, \ie\ the integral of the
matched PDF is different from the integral of the input quasi-PDF, and not
necessarily equal to 1 over the whole integration range. As a remedy, a modified
matching expression, which is given explicitly in Eq.~\eqref{eq::matching} of
App.~\ref{app:coefficients}, was proposed in Ref.~\cite{Alexandrou:2018pbm}. It
consists in resorting to pure plus functions when subtracting the logarithmic
divergence in self-energy corrections. However, this is an additional
subtraction with respect to the minimal subtraction of the $\MSb$ scheme and
thus, defines a modified $\MSb$ scheme, the so-called $\MMSb$ scheme. As such,
it requires the quasi-PDF to be expressed in this modified scheme. The
expression for the conversion of $\MSb$-renormalized matrix elements to the
$\MMSb$ scheme was worked out in Ref.~\cite{Alexandrou:2019lfo} and we refer to
it for the details of the procedure. Nevertheless, this modification is
numerically very small, as also shown in Ref.~\cite{Alexandrou:2019lfo}. An
alternative modification of the $\MSb$ scheme that guarantees vector current
conservation was derived in an updated version of Ref.~\cite{Izubuchi:2018srq}.
This defines the so-called ``ratio'' scheme. In this scheme, only pure plus
functions are used, like in the $\MMSb$ scheme, but the modification is done
also for the ``physical'' region of $0<\xi<1$ (in the notation of
Eq.~\eqref{eq::matching}). Thus, the expected numerical effect of this
modification is larger, as shown explicitly in Ref.~\cite{Alexandrou:2019lfo}.
For this reason, we choose to use the $\MMSb$ procedure, with details of the
lattice computation of the bare matrix elements and the renormalization in the
$\MMSb$ scheme outlined in the next section. Yet another possibility of
performing the matching consists in directly relating the quasi-PDFs in the
intermediate RI-type scheme to $\MSb$ light-cone PDFs. This was proposed in
Ref.~\cite{Stewart:2017tvs} for the unpolarized case. Obviously, such
procedure is equivalent to the one adopted here, with possibly different
systematic effects. All of the discussed papers considered the matching to only
first order in perturbation theory. It remains an important direction to derive
the two-loop formulae, which would allow to estimate the size of perturbative
truncation effects in the conversion from the intermediate lattice scheme to
$\MSb$ and in the matching itself.

\section{Nonsinglet distributions from quasi-PDFs Matrix Elements}
\label{sec:PDFstoqPDFs}

In this section, we describe the lattice data used in this work, presenting
briefly the quasi-PDFs matrix elements (MEs) computed in
Refs.~\cite{Alexandrou:2018pbm, Alexandrou:2019lfo}. Using the results recalled
in Sec.~\ref{sec:theoryoverview}, we show that we can factorize such matrix
elements into two nonsinglet distributions and a perturbatively computable
coefficient, just as if they were experimental data for high-energy cross
sections.

\subsection{Lattice data}
\label{subsec:latticedata}
The field of nucleon isovector ($u-d$) quasi-PDFs has matured in recent years.
Exploratory studies for all types of collinear PDFs -- unpolarized, helicity and
transversity -- were performed in
2014-2016~\cite{Lin:2014zya,Alexandrou:2015rja,Chen:2016utp,Alexandrou:2016jqi}.
They used lattice ensembles with non-physical pion masses and the results had
unsubtracted divergences, due to the lack of a well-defined renormalization
procedure. The latter was proposed and applied for the first time in
Refs.~\cite{Constantinou:2017sej,Alexandrou:2017huk}, utilizing a variant of the
regularization-independent momentum subtraction scheme
(RI'-MOM)~\cite{Martinelli:1994ty}. Moreover, another major progress for
unpolarized PDFs was the identification of a lattice-induced mixing between the
bilinear operator used in the first exploratory studies, which was defined using
$\gamma_z$ to determine the Dirac structure, and the scalar bilinear operator
(in spin space)~\cite{Constantinou:2017sej}. Even though in principle it is
possible to compute the matrix elements of the latter and a mixing
renormalization matrix to properly subtract the
divergences~\cite{Alexandrou:2017huk}, this is bound to lead to much
deteriorated precision, due to the rather poor signal for the scalar operator.
Instead, it is preferable to define the quark bilinear using the $\gamma_0$
Dirac matrix, since this procedure does not give rise to mixing. Moreover the
quasi-PDF computed with it converges faster in powers of $1/P_z^2$ to the
light-cone PDF, as argued in Ref.~\cite{Radyushkin:2016hsy}. Summarising in just
one sentence, we could say that the major progresses with respect to the early
works for unpolarized quasi-PDFs came from: (1) change of the Dirac structure in
order to avoid the mixing, (2) non-perturbative renormalization procedure, (3)
simulations at the physical pion mass. Matrix elements corresponding to such
setup were computed in Refs.~\cite{Alexandrou:2018pbm,Alexandrou:2019lfo} and they are briefly
described below. For a recent review of other available results for quasi-PDF
matrix elements, see \eg\ Ref.~\cite{Cichy:2018mum}. 

\begin{figure}[h]
	\minipage{0.45\textwidth}
	\includegraphics[width=12cm,height=5cm,keepaspectratio]{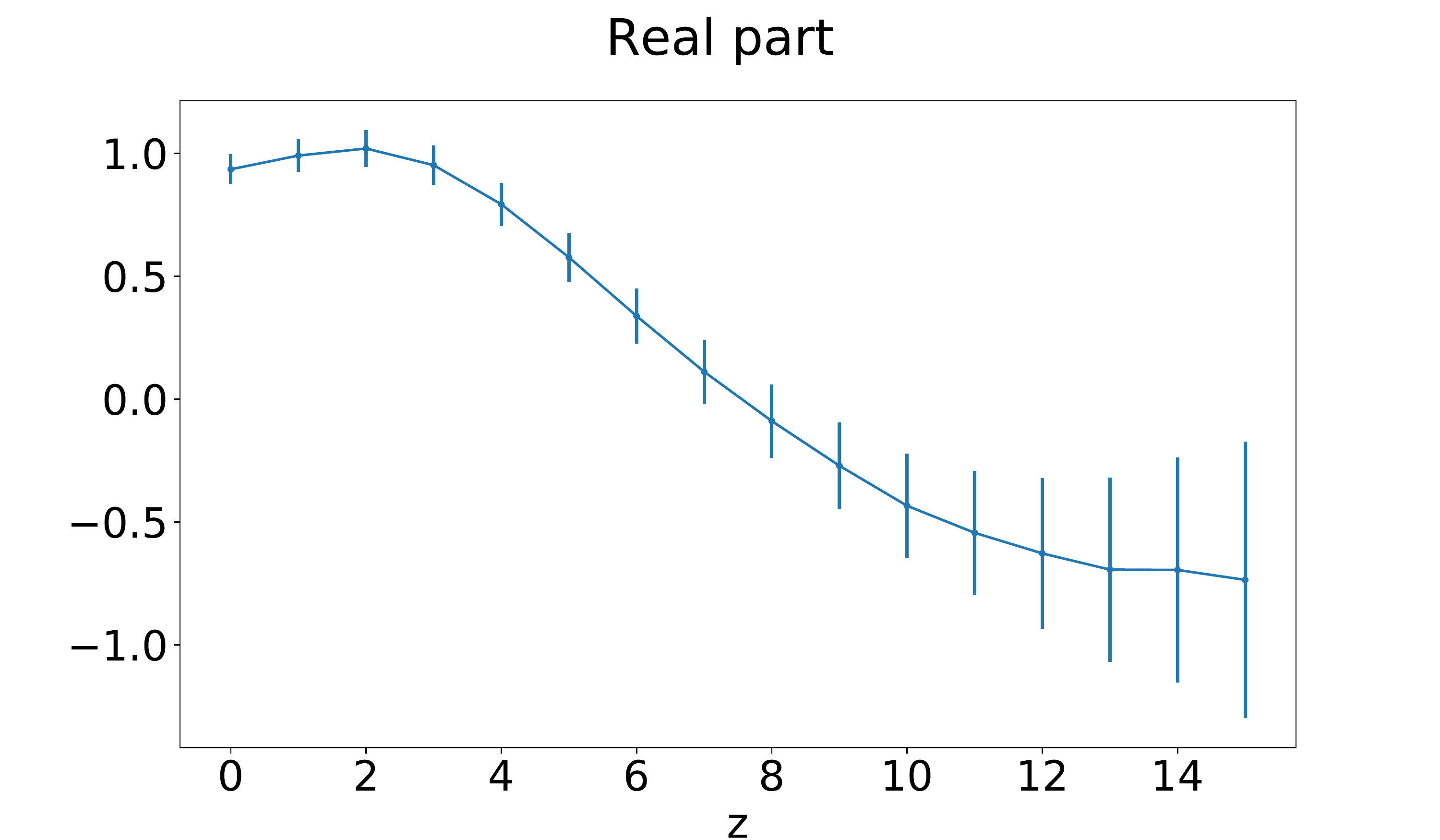}  
	\endminipage\hfill 
	\minipage{0.45\textwidth}
	\includegraphics[width=12cm,height=5cm,keepaspectratio]{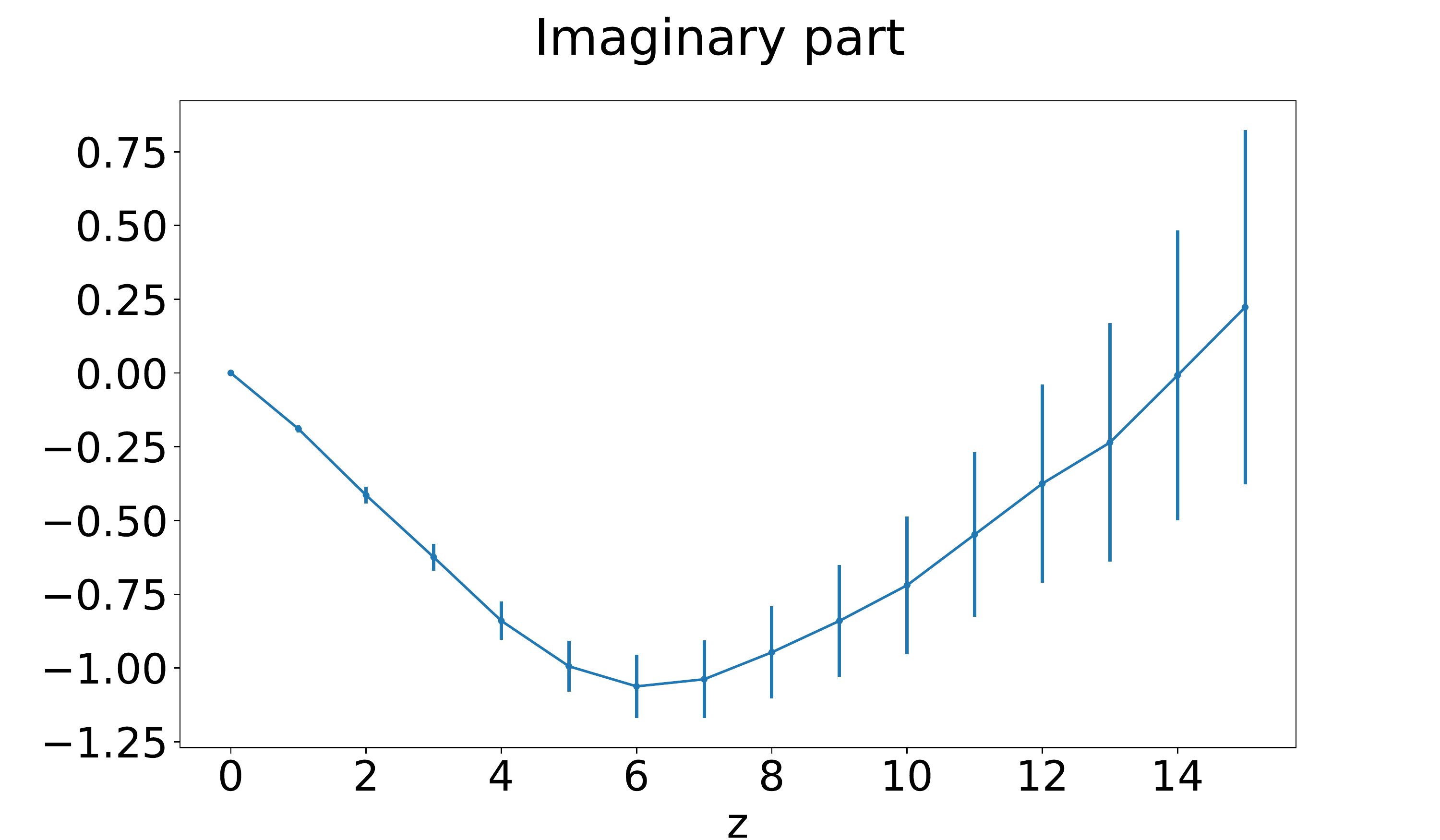}  
	\endminipage\hfill 
	\vspace*{-3mm}
	\caption{Real (left) and imaginary (right) part 
         of the quasi-PDF ME for the data used in this work, computed in Refs.~\cite{Alexandrou:2018pbm, Alexandrou:2019lfo}.
         The error band displayed accounts only for statistical uncertainty.}
	\label{fig::data}
\end{figure}

The data used in this work to illustrate the impact of lattice calculations on
phenomenological fits were computed by the Extended Twisted Mass Collaboration
(ETMC)\footnote{Until 2018 known as the European Twisted Mass Collaboration.}.
They used one ensemble of gauge field configurations with two degenerate light
quarks~\cite{Abdel-Rehim:2015pwa} with masses chosen to reproduce the physical
value of the pion mass ($m_\pi\approx130$ MeV, \ie\ slightly below the actual
physical value). The lattice spacing is $a=0.0938(3)(2)$~fm~\cite{Alexandrou:2017xwd} and the lattice has
$48^3 \times 96$ sites, corresponding to the spatial extent $L$ of around 4.5~fm
and $m_\pi L = 2.98$. ETMC calculated bare quasi-PDF matrix elements for the
unpolarized, helicity and transversity cases, but we concentrate only on the
unpolarized one. The lattice data are available for three nucleon boosts,
$P_z=6\pi/L,\,8\pi/L$ and $10\pi/L$ (0.83 GeV, 1.11 GeV and 1.38 GeV in physical
units) and for four values of the temporal separation between the nucleon
creation and annihilation operators, $t_s/a{=}8,9,10,12$ ($0.75,\, 0.84,\,
0.94,\, 1.13$~fm). As shown in Refs.~\cite{Alexandrou:2018pbm,Alexandrou:2019lfo}, there are signs
of convergence in the nucleon momentum (the largest two momenta give compatible
results), indicating that the boost is already enough to suppress higher-twist
effects below statistical precision. Moreover, as pointed out in
Ref.~\cite{Alexandrou:2019lfo}, excited-states contamination at the largest
source-sink separation is small, \ie\ the single-state fits at this $t_s$ are
compatible with two-state fits including all four values of $t_s$. Hence, for
the purpose of this study, we consider only the data at the largest nucleon
boost and at the largest source-sink separation. 
They are shown in Fig.~\ref{fig::data}.

The bare lattice data contain two types of divergences. First of all, there are
standard logarithmic divergences with respect to the regulator, \ie\ terms that
behave like $\log(a\mu)$. Additionally, for non-zero Wilson line lengths,
further power-like divergences appear. They resum into a multiplicative
exponential factor, $e^{\delta m |z|/a}$, where $\delta m$ is
operator-independent. The desired renormalization scheme for the final results
is the $\MSb$ scheme of dimensional regularization. However, obviously, the
latter is impossible on a lattice, restricted to integer dimensions. Thus, the
usage of an intermediate lattice renormalization scheme is required. In
Ref.~\cite{Alexandrou:2017huk}, it was proposed to use an RI'-type prescription.
The renormalization conditions are enforced on the amputated vertex functions of
operators with different Wilson line lengths $z$, setting them to their
tree-level values. A similar renormalization condition is applied for the quark
propagator. This results in a set of matrix elements renormalized in the RI'
scheme. Thus, a perturbative conversion from the RI' to the $\MSb$ scheme is
needed. Such a conversion was derived in Ref.~\cite{Constantinou:2017sej} to
one-loop order and was applied to the RI'-renormalized matrix elements. As we
discussed in the previous section, to guarantee vector current conservation, we
use a modified $\MSb$ scheme, termed the $\MMSb$ scheme. Thus, another
perturbative conversion of the $\MSb$-renormalized matrix elements is required,
according to the formula given in Ref.~\cite{Alexandrou:2019lfo}. After this
conversion, renormalized matrix elements in the $\MMSb$ scheme are the starting
point of the current analysis.

It is important to emphasize that despite having numerical evidence for the
smallness of the effects of the nucleon momentum and of excited states, matrix
elements from lattice studies come with a variety of other systematic effects.
We discuss them in the next subsection. For more details about the lattice
computation of the matrix elements, we refer the reader to
Ref.~\cite{Alexandrou:2019lfo}.

\subsection{Systematics in matrix elements of quasi-PDFs}
\label{subsec:sys}

A proper investigation of systematic effects in matrix elements evaluated in
lattice QCD simulations is a difficult task, necessitating dedicated efforts.
Such efforts consist in simulating with varied parameter values, such as the
lattice spacing, the lattice volume, or the temporal separation between the
source and the sink in nucleon three-point functions. Moreover, unrelated to the
lattice regularization, there are theoretical uncertainties intrinsic to the
quasi-distribution approach whose impact should also be assessed~\footnote{Note
that theoretical uncertainties can be included in global fits of PDFs as
detailed in Ref.~\cite{AbdulKhalek:2019bux, AbdulKhalek:2019ihb}}. For an extensive review of these
different uncertainties, we refer to Refs.~\cite{Cichy:2018mum,Monahan:2018euv},
while a discussion of the systematic effects in the ETMC quasi-PDFs computation
can be found in Ref.~\cite{Alexandrou:2019lfo}. The latter contains the analysis
of the effects investigated so far and a discussion of directions that need to
be pursued to fully quantify all the relevant systematics.

Here, we briefly summarize the conclusions reached up to the present stage. It
is important to emphasize that while the impact of some systematics is already
known to a reasonable degree, reliable estimates of certain types of effects are
still largely unknown. Nevertheless, rough assessments can be made even in the
case of missing lattice data, by looking at the behaviour of related observables
such as the average quark momentum fraction or nucleon charges that have a long
history of evaluations on the
lattice~\cite{Syritsyn:2014saa,Constantinou:2014tga,Constantinou:2015agp,Alexandrou:2015xts,Green:2018vxw}.
This allows us to build scenarios describing the potential impact of the
systematics on the matrix elements of quasi-PDFs. We consider three scenarios
where the systematic effect is a given percentage of the central value of the
matrix element and three further ones where it is a given additive shift. We
always exclude from the analysis the imaginary part of the matrix element at
$z=0$, equal to 0 by antisymmetry with respect to the sign change of $z$.

\textbf{Cut-off effects}. One of the most obvious systematic effects in lattice
computations comes from the finite value of the lattice spacing, $a$, \ie\ the
ultraviolet cut-off imposed for the regularization of the theory. While a proper
investigation of this uncertainty requires explicit simulations at a few values
of the lattice spacing, which are still missing for quasi-PDFs, we may assume
that discretization effects are not excessive. This expectation is based on two
indirect, but related premises. First, one of the manifestations of large
cut-off effects is the violation of the continuum relativistic dispersion
relation, which is, however, not observed in the lattice data in
Ref.~\cite{Alexandrou:2019lfo}. Second, the first moment of the unpolarized
$u-d$ PDF gives the quark momentum fraction $\langle x\rangle_{u-d}$. This
quantity was intensively investigated on the lattice and we may take the typical
size of discretization effects found in such studies. Looking at a summary plot
including data from different lattice groups, such as Fig.\ 12 from Ref.\
\cite{Constantinou:2014tga}, we see that cut-off effects at lattice spacings
comparable to the one of the present work are typically at the 5-15\% level in a
fixed lattice setup (same discretization, pion mass, volume etc.). Thus, we
investigate 6 plausible choices for the magnitude of cut-off effects: 10\%,
20\%, 30\% of the matrix element and additive effects of 0.1, 0.2 and 0.3. 

\textbf{Finite volume effects (FVE)}. Another natural source of uncertainty in
all lattice simulations is the finite size of the box, $L$, which acts as an
infrared regularization. Similarly to discretization effects, a robust
investigation of these effects necessitates running the computations for a few
values of the lattice size. However, the difference with respect to the lattice
spacing effects, typically linear in $a$ or $a^2$ in the asymptotic scaling
regime, is that FVE are usually suppressed as $\exp(-m_\pi L)$, where $m_\pi$ is
the pion mass. This leads to typically $\mathcal{O}(1-5\%)$ effects in hadron
structure observables if $m_\pi L\geq 3$. For the matrix elements used in this
work, $m_\pi L\approx3$ -- thus, the reasonable assumption about the size of FVE
is approx.\ 5\%. In addition to these ``standard'' FVE of lattice computations,
it has been recently pointed out that the usage of a spatially extended
operator, including a Wilson line, may lead to additional
FVE~\cite{Briceno:2018lfj}. The intuition behind this is that further FVE may
appear when the Wilson line has non-negligible size with respect to the lattice
length in the boost direction. The analysis of Ref.~\cite{Briceno:2018lfj}
pertains to a toy scalar theory and predicts a FVE of the form $\exp(-M (L-z))$
(possibly with a polynomial amplifying prefactor), with $M$ being the analogue
of the mass of the investigated hadron in the quasi-PDF approach. Given that the
nucleon mass is at the physical point around 7 times larger than the pion mass,
that would lead to totally irrelevant effects, since the maximum considered $z$
is more than 3 times smaller than $L$. However, it can not be excluded that in
QCD, the form of this FVE can be more severe, e.g.\ $\exp(-m_\pi (L-z))$. With
the physical $m_\pi$ and $z_{\rm max}\approx L/3$, this could lead to the
amplification of FVE from $\mathcal{O}(5\%)$ to even above 10\% at large $z$. We
remark that ETMC has investigated FVE in the renormalization functions for the
matrix elements and found no sign of excessive FVE coming at large $z$ (total
FVE not larger than around 3\%)~\cite{Alexandrou:2019lfo}. We investigate 3
scenarios for fixed percentage effects: constant FVE of 2.5\% and 5\%, as well
as $z$-dependent ones of the form $\exp(-3+0.062z/a)\%$, where 0.062 is the pion
mass value for the present ensemble, expressed in lattice units. In addition, we
consider 3 shifts: 0.025, 0.05 and $\exp(-3+0.062z/a)$.

\textbf{Excited states contamination}. One of the key uncertainties in nucleon
structure calculations is whether the ground state hadron state is isolated. If
the temporal separation between the interpolating operators creating the nucleon
and annihilating it is too small, uncontrolled excited states contamination may
appear, leading to a bias in the results. In the context of quasi-PDFs, an
important aspect is that this contamination strongly depends on the boost,
causing a delicate interplay between the need of large momentum, required for
robust matching to light-cone distributions, and excited states contamination,
larger for high boost. Thus, a careful analysis is needed to ensure ground state
dominance. Such an analysis was performed for the matrix elements used in this
work~\cite{Alexandrou:2019lfo}. The conclusion that we use for the present case
is that these matrix elements are safe against excited states effects at the
level of their statistical precision. In this way, we choose three values of
uncertainty from excited states: 5\%, 10\% and 15\%. When the renormalized
matrix elements are close to zero, the relative uncertainty can be larger and
thus, we consider also three additive scenarios with magnitude 0.05, 0.1 and
0.15.

\textbf{Truncation effects}. The perturbative ingredients of the quasi-PDF
approach are of two kinds. One of them is related to the fact that the lattice
approach works in integer dimensions and thus, dimensional regularization of the
$\MSb$ scheme is impossible. Instead, as discussed above, a non-perturbative
renormalization programme has been proposed by ETMC~\cite{Alexandrou:2017huk},
utilizing a variant of the regularization-independent momentum subtraction
scheme (RI'-MOM). The renormalization correlators obtianed in this way can then
be translated perturbatively to the $\MSb$ scheme and finally to the $\MMSb$
scheme, using formulae derived in Refs.~\cite{Constantinou:2017sej,
Alexandrou:2019lfo}. These formulae are currently available to the one-loop
level and thus subject to a truncation effect from unknown higher orders. A
manifestation of this effect is the fact that the $Z$-factors have a
non-vanishing imaginary part even after conversion to $\MSb$, where they should
be purely real. To evaluate the impact of this uncertainty, we compare the
renormalized matrix elements with the ones obtained from applying only the real
parts of the $Z$-factors. We find that the matrix elements obtained by this
alternative procedure are compatible with the actual ones within statistical
uncertainties, with relatively larger effects observed for small $z/a$ in the
imaginary part (up to $\mathcal{O}(5\%)$) and intermediate $z/a$ in the real
part (the real part is small there -- thus, the observed absolute effects of
around 0.2 can be a large percentage of the value). Apart from the scheme
conversion truncation effects, the necessary perturbative ingredient of the
approach is the matching between quasi-PDFs and light-cone distributions, also
known to one loop~\cite{Xiong:2013bka,Izubuchi:2018srq,Alexandrou:2019lfo}.
Without knowing the two-loop formulae, it is difficult to estimate their size.
Comparing the quasi-distribution and the resulting light-cone PDF, the numerical
magnitude of the matching factor can be significant and thus, the higher order
effects may be sizable. The ``natural'' size of such truncation effects is of
$\mathcal{O}(\alpha_s^2)$, which amounts to around 10\% at the renormalization
scale we consider. However, they are rather uncertainties of the procedure, so
they can not be translated to uncertainties of the matrix elements. These
uncertainties are the analogue of the theoretical uncertainties that come from a
truncated perturbative expansion in the description of observables in
phenomenological fits of the PDFs, and can be treated as mentioned in the
footnote above. Finally, we consider 6 scenarios for truncation effects
pertinent to matrix elements (\ie\ originating from the perturbative uncertainty
in $Z$-factors): 10\%, 20\%, 30\% of the central value of the matrix element, as
well as shifts of 0.1, 0.2 and 0.3. 

\textbf{Higher twist effects}. For the current analysis we decide to ignore the
effect of higher twists, \ie\ the presence of power-like corrections to the
factorization formula. At this preliminary stage, we are not concerned by their
effects, but a more precise phenomenological analysis should definitely take
those into account. We will come back to this point in the conclusions. 

\textbf{Other effects}. Apart from the systematics mentioned above, there are
some other effects that potentially affect the results. One of them is the usage
of a setup including two degenerate light quarks. However, this effect, working
in the isospin limit instead of taking into account the different masses and
electric charges of the light quarks, is expected to be much below the level of
the current precision -- of the order of the proton-neutron mass splitting, \ie\
at the per mille level. A similar magnitude can be expected for the contribution
of the neglected sea quark loops from heavier quarks. Such effects can at
present be safely ignored and will become important only when aiming at an
$\mathcal{O}(1\%)$ precision or better.

\begin{table}[h]
\begin{center}
\begin{tabular}{|c|c|c|c|c|}
\hline
Scenario & Cut-off & FVE & Excited states & Truncation\\
\hline
S1 & 10\% & 2.5\% & 5\%  & 10\%\\
S2 & 20\% & 5\%  & 10\%  & 20\%\\
S3 & 30\% & $e^{-3+0.062z/a}\%$  & 15\%  & 30\% \\
S4 & 0.1  & 0.025 & 0.05 & 0.1\\
S5 & 0.2  & 0.05  & 0.1 & 0.2\\
S6 & 0.3  & $e^{-3+0.062z/a}$ & 0.15 & 0.3\\
\hline%
\end{tabular}
\end{center}
\caption{\label{tab:systematics} Scenarios of the impact of different systematic effects in the renormalized matrix elements of quasi-PDFs. Percentage values for scenarios S1-S3 should be understood as a given fraction of the central value of the matrix element, while absolute values for S4-S6 are shifts independent from the matrix element.}
\end{table} 

\textbf{Final scenarios}. In the end, we define 6 scenarios of possible impact
of systematic effects, summarized in Tab.~\ref{tab:systematics}. Scenarios S1-S3
include uncer\-tain\-ties that are a fixed percentage of the central value of
the matrix element, while for S4-S6, the uncertainties are additive shifts
independent from the value of the matrix element. Scenarios S1, S4 can be
considered as the most ``optimistic'' ones. More realistic estimates of
uncertainties are included in S2 and S5. Finally, S3 and S6 are ``pessimistic'',
\ie\ assume largest plausible estimates of the various systematic effects.

\subsection{From parton distributions to lattice observables}
\label{subsec:thpredictions}
In this work, we aim at describing the data presented in the previous subsection;
further studies with more data and a more detailed treatment of systematic
errors are deferred to future publications. Hence, we specialize our discussion
to the case of the unpolarized isovector parton distribution. Following the
notations introduced above, the parton distribution $f_{3}$ is
defined as
\begin{align}
	f_{3}\lp x, \mu^2 \rp = 
	\begin{cases}
      \phantom{-} u\lp x, \mu^2 \rp - d\lp x, \mu^2\rp\, , \quad                 
      &\text{if} \,\,\,\,\, x > 0 \\
      -\bar{u}\lp -x, \mu^2 \rp + \bar{d}\lp -x, \mu^2 \rp\, , \quad 
      &\text{if} \,\,\,\,\, x < 0 
	\end{cases}
\end{align} 
where the support is given by $x\in \left[ -1, 1 \right]$. The factorization
theorem in Eq.~\eqref{eq::pdftoqpdf} becomes
\begin{align}
	\label{eq::factorization}
	  & \tilde{f}_{3}\left(x,\mu,P_z\right) = \int_{-1}^{+1} \frac{dy}{|y|}\,C_3\left(\frac{x}{y},\frac{\mu}{|y|P_z}\right)\, 
	  f_{3}\left(y,\mu^2\right)\, ,
\end{align}
where the quasi-PDF is the one given by $\Gamma = \gamma^0$ and the explicit
expression of the matching coefficients is given in
App.~\ref{app:coefficients}. Starting from the definition of quasi-PDFs
given in Eq.~\eqref{eq::bareqpdf}, it is clear that the lattice ME is given by the
inverse Fourier transform of Eq.~\eqref{eq::factorization}, which yields an
equation relating the light-cone PDFs on the right hand side to the lattice observable:
\begin{align}
	\label{eq:factorisedME}
	\text{h}_{\gamma^0,3}\lp z P_z, z^2,\mu\rp = 
	\int_{-\infty}^{\infty} dx \, e^{-i \lp x P_z \rp z} 
	\int_{-1}^{+1} \frac{dy}{|y|}\,
	C_3\left(\frac{x}{y},\frac{\mu}{|y|P_z}\right)\, 
	f_{3}\left(y,\mu^2\right). 
\end{align}
We can split the above complex identity into two real equations, relating the
real and imaginary part of the ME $\text{h}_{\gamma^0,3}\lp z \rp$ to the
light-cone distribution $f_{3}$. For the purpose of this work, we introduce two
lattice observables, denoted by $\mathcal{O}_{\gamma^0}^{\text{Re}}\lp z, \mu
\rp$ and $\mathcal{O}_{\gamma^0}^{\text{Im}}\lp z, \mu \rp$, defined as 
\begin{align}
	  & \mathcal{O}_{\gamma^0}^{\text{Re}}\lp z, \mu \rp 
		  \equiv \text{Re} \left[\text{h}_{\gamma^0,3}\lp z P_z, z^2, \mu^2 \rp\right] 
		  = \int_{-\infty}^{\infty} dx\, \cos{\lp x P_z z \rp} 
		  \int_{-1}^{+1} \frac{dy}{|y|}\, C_3\left(\frac{x}{y},\frac{\mu}{|y|P_z}\right)\, 
		  f_{3}\left(y,\mu^2\right), \\
	  & \mathcal{O}_{\gamma^0}^{\text{Im}}\lp z, \mu \rp 
		  \equiv \text{Im} \left[\text{h}_{\gamma^0,3}\lp z P_z, z^2, \mu^2 \rp\right] 
		  = \int_{-\infty}^{\infty} dx\, \sin{\lp x P_z z \rp} 
		  \int_{-1}^{+1} \frac{dy}{|y|}\, C_3\left(\frac{x}{y},\frac{\mu}{|y|P_z}\right)\, 
		  f_3\left(y,\mu^2\right)\, , 
\end{align}  
where we have only included $z$ and $\mu$ in the arguments of
$\mathcal{O}_{\gamma^0}^{\text{Re}}$ and $\mathcal{O}_{\gamma^0}^{\text{Im}}$ in
order to simplify the notation -- since we are working here with only one value
of $P_z$ there is little advantage in keeping all the arguments. The explicit
expression of $C_3$ contains plus distributions. Making them explicit we can
write the equations above as
\begin{align}
	\label{eq::V3factorization}
	  & \mathcal{O}_{\gamma^0}^{\text{Re}}\lp z, \mu \rp = \int_{0}^{1} dx \,\, \mathcal{C}_3^{\text{Re}}\lp x, z, \frac{\mu}{P_z}  \rp V_3\left(x,\mu\right) = \mathcal{C}_3^{\text{Re}}\lp z, \frac{\mu}{P_z}  \rp \circledast V_3\left(\mu^2\right), \\
	\label{eq::T3factorization}
	  & \mathcal{O}_{\gamma^0}^{\text{Im}}\lp z, \mu \rp = \int_{0}^{1} dx \,\, \mathcal{C}_3^{\text{Im}}\lp x, z, \frac{\mu}{P_z}  \rp T_3\left(x,\mu\right)= \mathcal{C}_3^{\text{Im}}\lp z, \frac{\mu}{P_z}  \rp \circledast T_3\left(\mu^2\right)   
\end{align}
where $V_3$ and $T_3$ are the nonsinglet distributions defined by
\begin{align}
	  & V_3 \lp x \rp = u\left(x\right) - \bar{u}\left(x\right) -\left[d\left(x\right)-\bar{d}\left(x\right)\right]\, , \\
	  & T_3 \lp x \rp = u\left(x\right) + \bar{u}\left(x\right) -\left[d\left(x\right)+\bar{d}\left(x\right)\right]\, ,
\end{align}
where, for simplicity, the $\mu$ dependence has been omitted. 
The equations above relate the position space matrix elements computable on the lattice with
the collinear PDFs. Similar expressions were worked out in Ref.~\cite{Bali:2017gfr} in the context of
the pion distribution amplitude.   
The proof of Eqs.~\eqref{eq::V3factorization}, \eqref{eq::T3factorization} does require some
care, and it is fully worked out in App.~\ref{app:coefficients}, together with
the explicit expressions for the coefficents
$\mathcal{C}_3^{\text{Re},\text{Im}}$. A discussion about the convergence of the
integrals involved is also reported there. The above results show how fixed $z$
matrix elements defining the quasi-PDF in position space give direct access to
two indipendent nonsinglet distributions, through the integration of the parton
distribution over its full support with a perturbatively computable coefficient.
We will denote this operation as $\circledast$. 

It is useful at this point to recall the form of the QCD factorization formula
for the DIS nonsinglet structure function. Considering $F_2^{\text{NS}}\lp x,Q^2
\rp \equiv F_2^p\lp x,Q^2 \rp - F_2^d\lp x,Q^2 \rp $, where $F_2^p$ and $F_2^d$
are the structure functions of the proton and of the deuteron respectively, we
have
\begin{align}
	\label{eq::DISfactorization}                                                F_2^{\text{NS}}\lp x,Q^2 \rp = \int_x^1 \frac{dy}{y}\,\mathcal{C}^{\text{DIS}}_3\lp \frac{x}{y}, \frac{Q^2}{\mu^2}, \as \rp T_3 \lp y, \mu^2 \rp = \mathcal{C}^{\text{DIS}}_3\lp\frac{Q^2}{\mu^2}, \as \rp \otimes T_3 \lp\mu^2 \rp 
\end{align}
Comparing Eqs.~\eqref{eq::V3factorization}, \eqref{eq::T3factorization}  with
Eq.~\eqref{eq::DISfactorization}, we see that the lattice observables introduced
above can be treated on the same footing as experimental data for DIS structure
functions, as they are related to the nonsinglet distributions through a
convolution with a coefficient that can be computed in perturbation theory.
However, the form of such convolution, denoted by $\circledast$ , is quite
different from the one appearing in the DIS case, denoted by $\otimes$: the
former involves a DIS-like convolution first, to go from the PDFs to quasi-PDFs,
followed by an integration over the full $x$-range to go to position space. This
suggests that this kind of convolution, if implemented in a PDFs fit, may
constrain the output much more than what the standard DIS convolution can do.

\section{Neural network fits}
\label{sec:fit}
In this section, we set up a neural net fit based on the results presented in
Sec.~\ref{sec:PDFstoqPDFs}. The implementation is done within the {\tt NNPDF}
framework. We begin by briefly recalling the main features of such fits,
focusing on the parametrization of the parton distributions, the minimization
and cross-validation algorithms and the Monte Carlo replicas approach, referring
to the specific {\tt NNPDF} publications for more details.
We then describe in detail the implementation of
Eqs.~\eqref{eq::V3factorization}, \eqref{eq::T3factorization}, and in particular
the construction of FastKernel tables for the lattice observables. Once the
observables can be computed using a FastKernel table, their inclusion in a PDF
fit within the {\tt NNPDF} environment becomes straightforward.

\subsection{Fitting strategy and general settings}
\label{subsec:nnpdf}
Given an ensemble of data whose values can be computed as a convolution of a
perturbative Wilson coefficient and the PDF using some factorization theorem,
the PDFs can be extracted from a minimum-$\chi^2$ fit to these data. It is
worthwhile to stress once again that in this respect there is no difference
between the lattice data and experimental data. In order to define the $\chi^2$
a parametrization of the functional form is required; the minimization is then
performed as a function of the parameters that define such functional form. In
our case, and more generally in the {\tt NNPDF} approach, the functional form of
the PDFs is  given by a single-layer neural network, and the parameters are the
weights and the biases of the neural network~\cite{DelDebbio:2007ee}. The parton
distributions independentely parametrized are the gluon and the singlet
distribution, $\lp g,\,\Sigma \rp$, which mix under evolution, and the
nonsinglet distributions given by $\lp V, T_3, V_3,
T_8, V_8, T_{15} \rp$, whose definition is given for
example in Ref.~\cite{Ball:2008by}.
As discussed in the previous sections, the lattice data we consider here give
access only to $T_3$ and $V_3$. Since the mixing with other
flavors does not occur neither in the matching nor in the DGLAP evolution, our
fits yield results for these two distributions only. The parametrization of
their $x$-dependence at the fitting scale $\mu_0 = 1.65$ GeV is 
\begin{align}
    V_3 \lp x, \mu_0^2\rp \propto x^{\alpha_V}\lp 1- x \rp ^{\beta_V} 
    \text{NN}_V\lp x \rp \, ,\\
    T_3 \lp x, \mu_0^2\rp \propto x^{\alpha_T}\lp 1- x \rp ^{\beta_T} 
    \text{NN}_T\lp x \rp \, ,
\end{align}
where NN denotes a neural net, and the parameters $\alpha$ and $\beta$ are
additional free parameters, which will be fitted during the training. As
extensively discussed in several {\tt NNPDF} publications, see for example Refs.~\cite{Ball:2008by, Ball:2010de}, the function
multiplying the neural net improves the stability and the speed of the
minimization procedure, without introducing a bias in the result.

Having chosen a parametrization for the PDFs, the optimal fit is determined by
varying its free parameters in order to minimize some figure of merit,
representing the fit quality. This is given by the $\chi^2$ function, defined as
\begin{align}
	\label{eq::chi2}                   
	\chi^2\left(\theta\right) = \frac{1}{N_{\text{dat}}}\sum_{i,j} \left(\mathcal{O}\lp z_i \rp-\mathcal{O}^{\text{th}}\lp z_i, \theta \rp \right)\left[\text{Cov}^{-1}\right]_{\text{ij}}\left(\mathcal{O}\lp z_j \rp-\mathcal{O}^{\text{th}}\lp z_j, \theta \rp \right), 
\end{align}
where $\mathcal{O}\lp z_i \rp $ denotes the measured lattice observable and
$\mathcal{O}^{\text{th}}\lp z_i, \theta \rp$ is the corresponding theoretical
prediction, in terms of the matching coefficient and the parametrized parton
distribution, given in Eqs.~\eqref{eq::V3factorization},
\eqref{eq::T3factorization}. We denote by $\theta$ the set of free parameters
entering the neural net and the preprocessing term. The covariance matrix
entering the $\chi^2$ definition is defined to take into account the
distribution of the experimental data entering the fit and their correlations.
It is used for both the $\chi^2$ definition and the generation of Monte Carlo
replicas, as described below. In the specific case of lattice data, it has to be
constructed using the information about the statistical and systematic
uncertainties that arise in a lattice QCD simulation. The different sources of
systematics are described in Sec.~\ref{subsec:sys}, and the scenarios we 
consider here are summarized in Tab.~\ref{tab:systematics}.

The minimization of the $\chi^2$ is performed through the CMA
algorithm~\cite{DBLP:journals/corr/Hansen16a, Bertone:2017tyb}, employing a cross-validation
technique to avoid overfitting. In this method, the available data are split in
two sets. The first, the training set, is used for the minimization of the error
function, while the second, the validation set, does not enter the fitting
proceedure. At each iteration of the minimization algorithm, the error function
between the theory predictions from the neural net and the data is computed for
both the training and validation set. At an early stage of the training, both
these quantities are expected to decrease. However, towards the end of the
training, while the error function over the training set will keep decreasing,
the same value computed over the validation data will reach a minimal value, and
eventually it may even start increasing. This is a signal of overfitting, and
the point in parameter space yielding the minimal value of the validation error
is the one taken as the fit result. It is important to notice that such
procedure is even more important when a small amount of data is available, like
in the present case, since the less data there are the easier it is to fit their
statistical noise.

In the {\tt NNPDF} framework a Monte Carlo approach is implemented in order to
get a reliable estimate of the PDF error. In this method an ensemble of
$N_{\text{rep}}$ artificial data is generated for each experimental point,
assuming a multigaussian distribution given by the experimental covariance
matrix. $N_{\text{rep}}$ independent fits are performed, generating a Monte
Carlo ensemble of PDFs that faithfully reproduces the statistical features of
the original experimental (or lattice) data. The Monte Carlo method therefore
propagates the error from the data to the PDFs set in a natural way, without the
need of any assumption on the way error propagation happens.

\subsection{FastKernel implementation}
\label{subsec:fk}

One of the primary issues in PDFs fits is the computational time required to
obtain the theoretical prediction for the experimental data entering the
definition of the $\chi^2$ in Eq.~\eqref{eq::chi2}: the parton distributions
have to be evolved from the fitting scale up to the observable scale, and then
they have to be convoluted with the correct coefficient function. As seen in
Sec.~\ref{subsec:thpredictions}, in the case of lattice observables the
integration of the parton distributions over their full support is needed. This
makes the form of the convolution $\circledast$ more complicated than the one we
usually have for other observables. Therefore, despite the fact that in this
work we will be using a small number of lattice data, we find it useful to
implement Eqs.~\eqref{eq::V3factorization}, \eqref{eq::T3factorization} by means
of FastKernel tables, introduced and validated in Refs.~\cite{Ball:2010de, Bertone:2016lga} in the
context of global QCD fits, and currently used within the {\tt NNPDF} code to
obtain all the required theoretical predictions in a global fit. From a
practical point of view, this also allows us to treat the lattice data on
exactly the same footing as the experimental ones, allowing a smooth and natural
way to introduce them in a parton distributions fit. In the following we briefly
revise this approach, referring to the original publication quoted above for
more details.

The lattice observables $\mathcal{O}_{\gamma^0}^{\text{Re},\text{Im}}\lp z,
\mu^2 \rp$ are determined at a given renormalization scale $\mu^2$. They can be
written in terms of the nonsinglet distributions at a given reference scale
$\mu_0^2$, by first evolving the parton distribution up to the scale $\mu^2$,
and then convoluting it with the coefficents
$\mathcal{C}_3^{\text{Re},\text{Im}}$ defined in
Eqs.~\eqref{eq::V3factorization}, \eqref{eq::T3factorization} and worked out in
App.~\ref{app:coefficients}. For the nonsinglet distributions considered in this work, the evolution is given by
\begin{align}
	\label{eq::NSevolutionPlus}
	T_3 \lp x, \mu^2 \rp = \int_x^1 \frac{dy}{y}\, 
	\text{K}_3^{(+)}\lp \frac{x}{y}, \as, \as^0 \rp T_3 \lp y, \mu_0^2 \rp\, , \\
	\label{eq::NSevolutionMinus}
	V_3 \lp x, \mu^2 \rp = \int_x^1 \frac{dy}{y}\, 
	\text{K}_3^{(-)}\lp \frac{x}{y}, \as, \as^0 \rp V_3 \lp y, \mu_0^2 \rp\, .
\end{align}
where the kernels $\text{K}^{(\pm)}$ are obtained by solving the DGLAP evolution
equations in the nonsinglet sector. For $\nsv$ and $\nst$ we have two different
nonsinglet evolution kernels, denoted by $\text{K}_3^{(-)}$ and
$\text{K}_3^{(+)}$ respectively. They are currently available from a number of
public codes, however one way they can be worked out and implemented is
summarised in the App.~\ref{app:dglap}. Eqs.~\eqref{eq::NSevolutionPlus}
and~\eqref{eq::NSevolutionMinus} can be rewritten expressing the parton
distribution in terms of an interpolation basis \cite{Ball:2010de}, for instance for the case of
$T_3$
\begin{align}
	  & T_3\lp x, \mu_0^2 \rp = \sum_{\beta}T_3\lp x_{\beta}, \mu_0^2\rp\mathcal{I}^{(\beta)}\lp x \rp 
          + \mathcal{O}\left[\left(x_{\beta+1}-x_{\beta}\right)^p\right], 
\end{align}  
where $p$ is the lowest order neglected in the interpolation.
In other words, the interpolating functions
act by picking up the value of the PDF at some point $x_{\beta}$ of a
predefined $x$-grid. Substituting in the evolution equation
Eq.~\eqref{eq::NSevolutionPlus} we get
\begin{align}
	\label{eq::fastevol}
	T_3 \lp x_{\alpha}, \mu^2 \rp & = 
	\sum_{\beta} \mathcal{K}^{(+)}_{\alpha\beta}\,
	T_3 \lp x_{\beta}, \mu_0^2 \rp\, . 
\end{align}
with
\begin{align}
	\mathcal{K}^{(+)}_{\alpha\beta} = \int_{x_{\alpha}}^1 \frac{dy}{y}\, 
	\text{K}^{(+)} \lp \frac{x_{\alpha}}{y}, \as, \as^0 \rp 
	\mathcal{I}^{(\beta)}\lp y \rp \, .
\end{align}
The interpolation basis used at the initial scale can also be used to
interpolate the parton distributions at the scale $\mu^2$ appearing in
Eqs.~\eqref{eq::V3factorization}, \eqref{eq::T3factorization}. For the imaginary part of the lattice observable we get
\begin{align}
	\label{eq::applgrid}             
	\mathcal{O}_{\gamma^0}^{\mathrm{Im}}\lp z,\mu \rp = 
	\sum_{\alpha} C_{z\alpha}^{\mathrm{Im}}\, 
	T_3\left(x_{\alpha},\mu^2\right)\, , 
\end{align}
with
\begin{align}
	C_{z\alpha}^{\mathrm{Im}} = \int_0^1 dx\, 
	\mathcal{C}_3^{\mathrm{Im}}\lp x, z, \frac{\mu}{P_z}\rp 
	\mathcal{I}^{(\alpha)}\lp x \rp\, . 
\end{align}
Putting together Eqs.~\eqref{eq::fastevol} and~\eqref{eq::applgrid} we get
\begin{align}
	\mathcal{O}_{\gamma^0}^{\mathrm{Im}}\lp z,\mu \rp = 
	\sum_{\beta}
	\mathcal{H}_{z\beta}^{\mathrm{Im}}\,
	T_3\lp x_{\beta}, \mu_0^2 \rp\, , 
\end{align}
where
\begin{align}
	\label{eq::FK}
	\mathcal{H}_{z\beta}^{\mathrm{Im}} = 
	\sum_{\alpha} C_{z\alpha}^{\mathrm{Im}}\, 
	\mathcal{K}^{(+)}_{\alpha\beta}\, . 
\end{align}
Eq.~\eqref{eq::FK} defines the FastKernel table which enters the computation of
the $\chi^2$ during the fit. It connects the parton distribution at the fitting
scale to the lattice observable, taking into account the QCD evolution, the
matching and the Fourier trasnform, expressing them through a single matrix
vector multiplication. Clearly a similar set of equations defines a FastKernel
table that yields the real part of the lattice observable,
$\mathcal{O}_{\gamma^0}^{\mathrm{Re}}$, as a function of the valence parton
distribution $V_3$.

\section{Results}
\label{sec:results}

Let us now proceed to presenting and discussing our numerical results. First, we
study the way the available lattice data might constrain the parton
distributions in a fit, by mean of closure tests: fake data for the real and
imaginary part of the ME are generated according to
Eqs.~\eqref{eq::V3factorization},~\eqref{eq::T3factorization} using as input a
chosen PDFs set. The fitting code is then run over these pseudo-data, using
exactly the same setting used in a common fit. By comparing the output of such
fits with the known input PDFs sets, we can assess the accuracy we may expect to
get from the current knowledge of the lattice data and their systematics. 

Then we present results for fits run over the data presented in
Sec.~\ref{subsec:latticedata}, studying the 6 different scenarios for the
treatment of the systematic errors described in Sec.~\ref{subsec:sys} and
summarized in Tab.~\ref{tab:systematics}. 
The results presented here have been produced using the {\tt NNPDF} fitting code \cite{Ball:2017nwa}
and the {\tt ReportEngine} software \cite{zahari_kassabov_2019_2571601}.  

\subsection{Closure tests}
\label{subsec:CT}
As shown in Sec.~\ref{subsec:thpredictions}, we can relate PDFs to lattice
observables through the matching convolution of Eq.~\eqref{eq::factorization}
followed by a Fourier transform. As already pointed out at the end of
Sec.~\ref{subsec:thpredictions}, the resulting convolution $\circledast$ is
quite different from the one entering standard QCD fits. In this section, we
assess how much this operation together with the available lattice data from
Refs.~\cite{Alexandrou:2018pbm,Alexandrou:2019lfo} are able to constrain parton distributions in a
fit, running some preliminary closure tests. For a detailed description of the
closure test procedure, we refer to Sec.~4 of Ref.~\cite{Ball:2014uwa}. We
generate pseudo-data corresponding to the data of Ref.~\cite{Alexandrou:2018pbm}
using {\tt NNPDF31\_nlo\_as\_0118} as our input PDFs set, and we run the fitting
code over them. The outcome of the closure test fit is then used to assess how
well the input PDFs can be reconstructed starting from the 16 position space ME
points and their uncertainties.

\begin{figure}[h]
	\minipage{0.45\textwidth}
	\includegraphics[width=12cm,height=5cm,keepaspectratio]{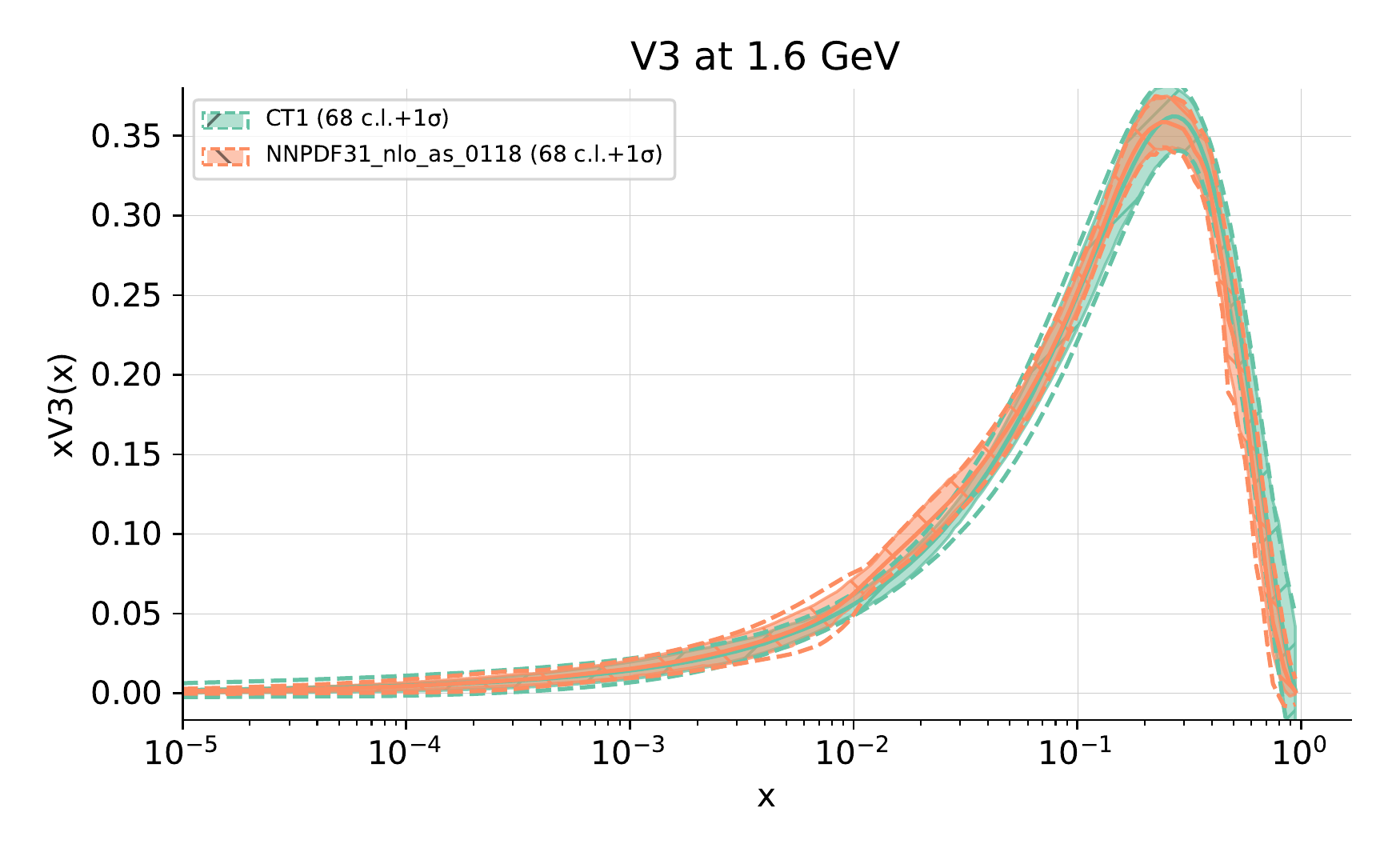}  
	\endminipage\hfill 
	\minipage{0.45\textwidth}
	\includegraphics[width=12cm,height=5cm,keepaspectratio]{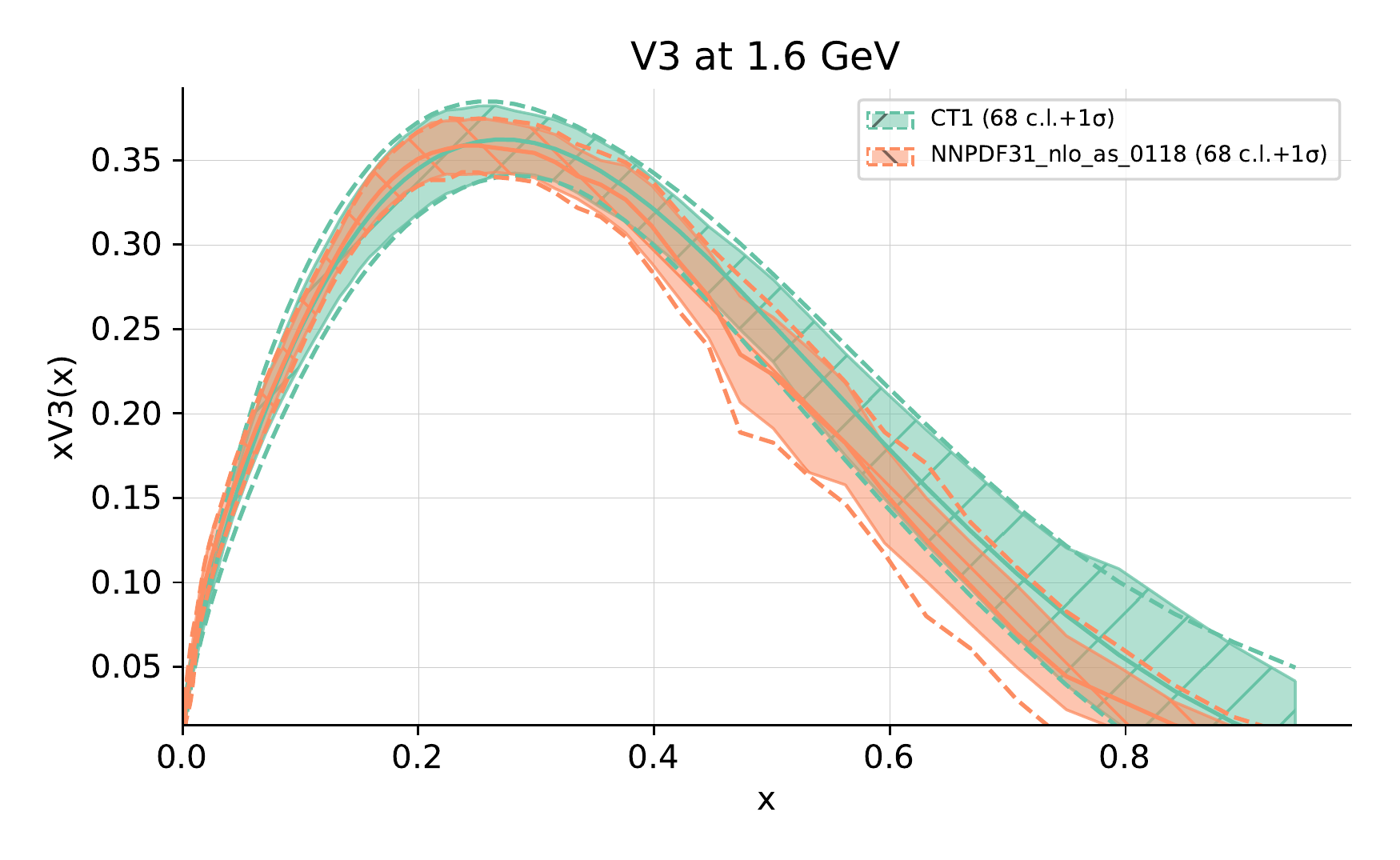}  
	\endminipage\hfill \\
	\minipage{0.45\textwidth} 
	\includegraphics[width=12cm,height=5cm,keepaspectratio]{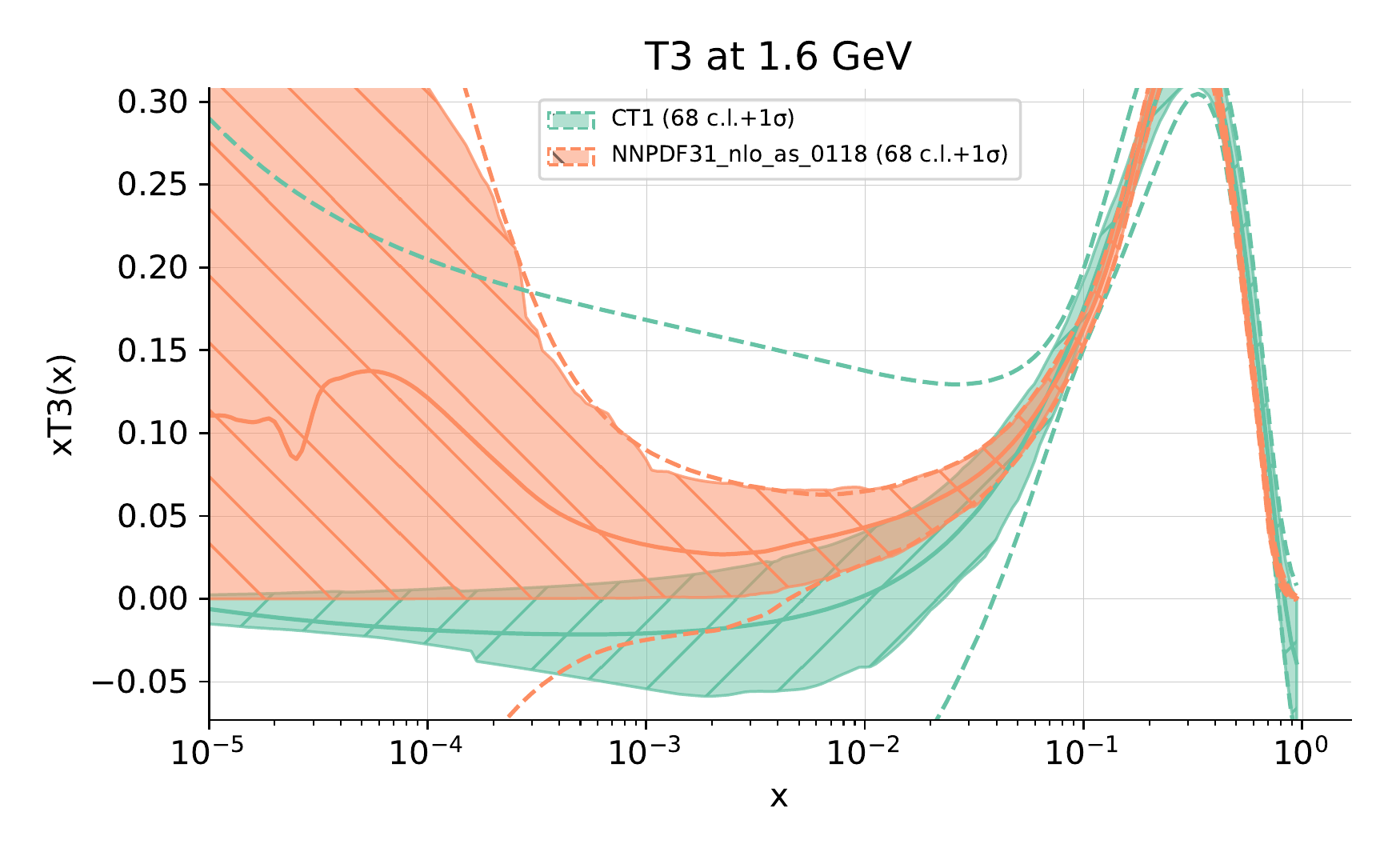}  
	\endminipage\hfill 
	\minipage{0.45\textwidth}
	\includegraphics[width=12cm,height=5cm,keepaspectratio]{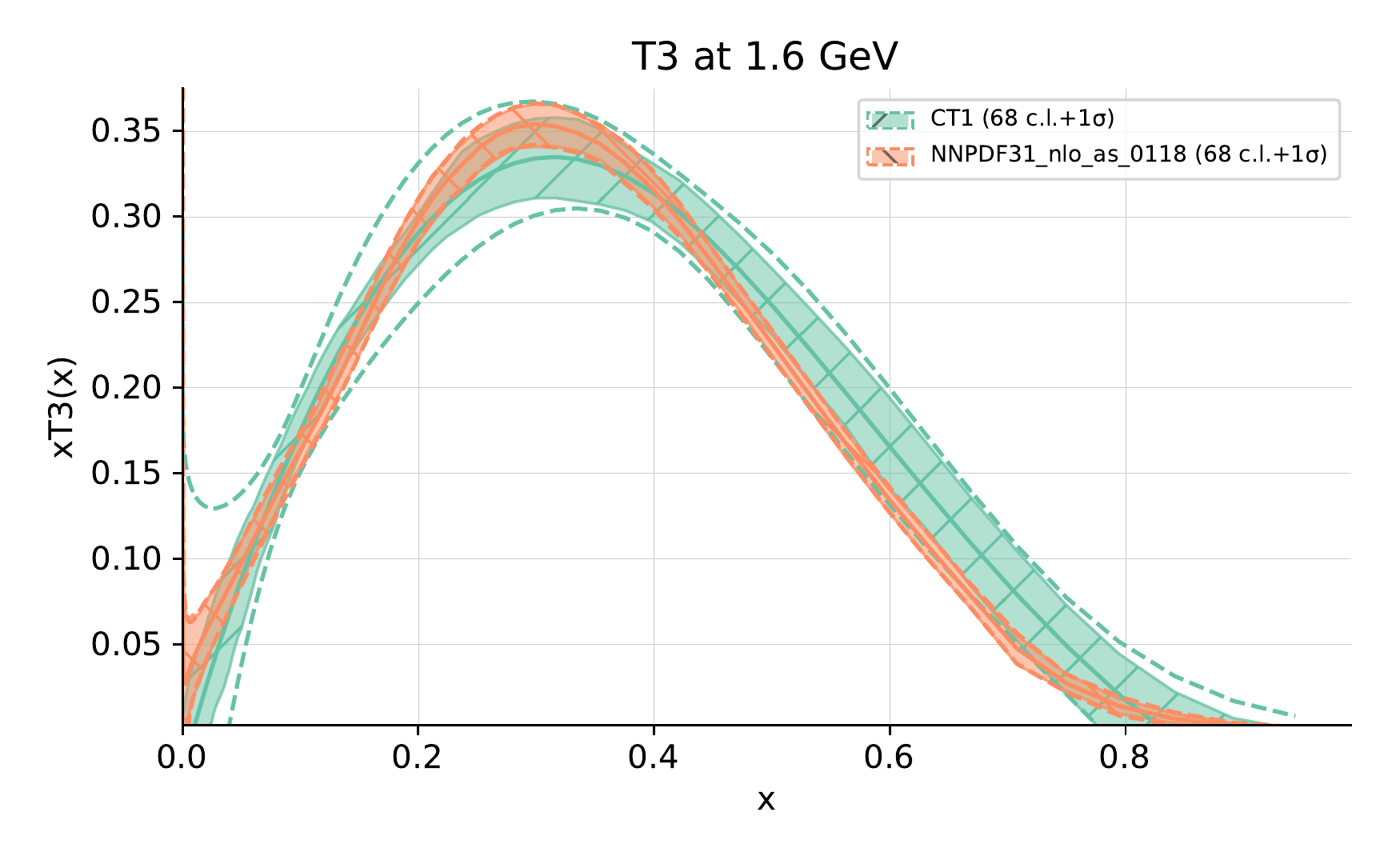}  
	\endminipage\hfill
	\vspace*{-5mm}
	\caption{Closure test fit with fixed small statistical error and no systematics (CT1) compared to the input PDFs set. 
		$V_3$ (top line) and $T_3$ (lower line) combinations in linear and logarithmic scale are shown.
	The input PDFs set is fully reconstructed within 1-sigma level, getting PDFs with an error band comparable to the input one.}
	\label{fig:1}
\end{figure}

\begin{figure}[h!]
	\minipage{0.45\textwidth}
	\includegraphics[width=12cm,height=5cm,keepaspectratio]{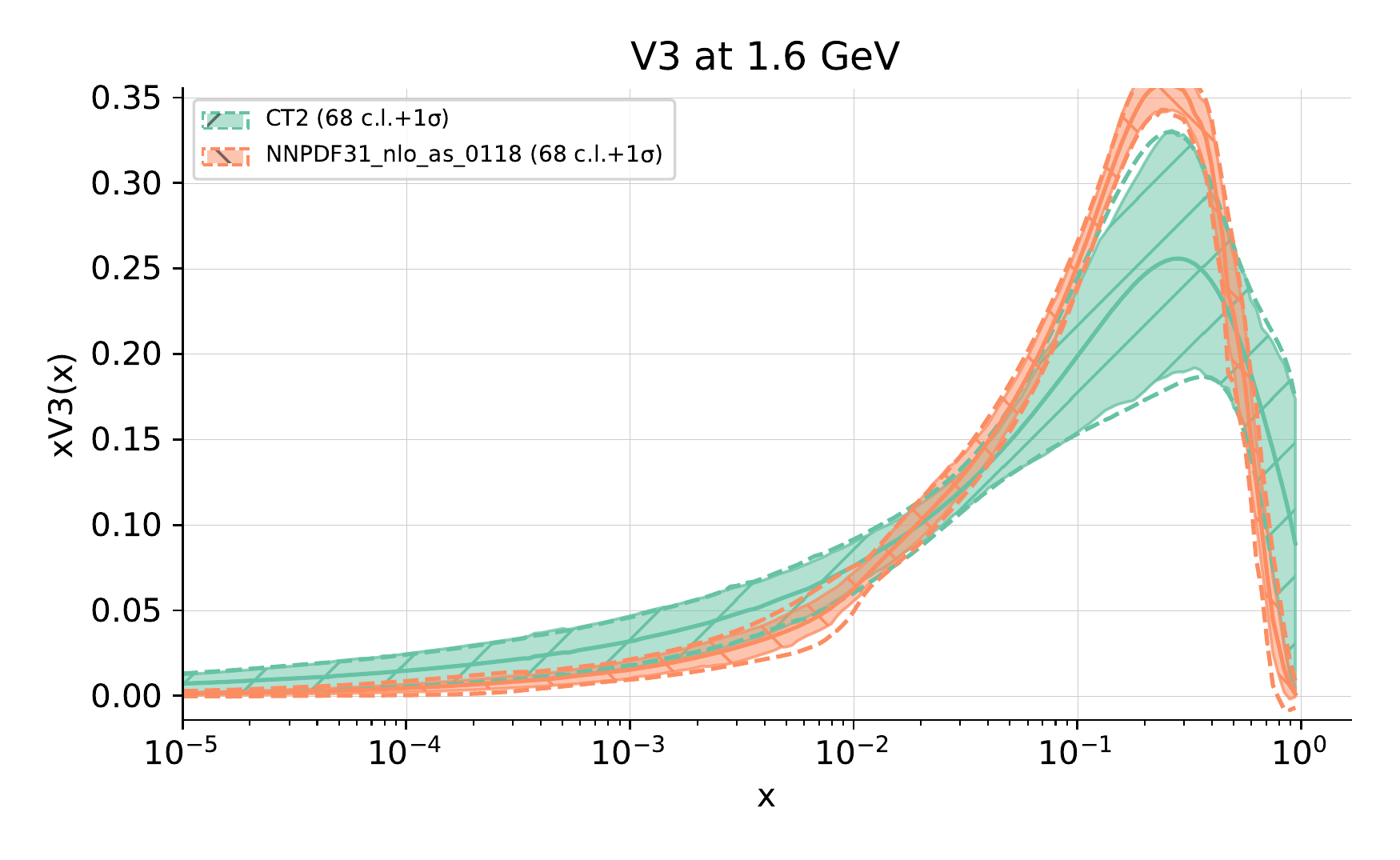}  
	\endminipage\hfill 
	\minipage{0.45\textwidth}
	\includegraphics[width=12cm,height=5cm,keepaspectratio]{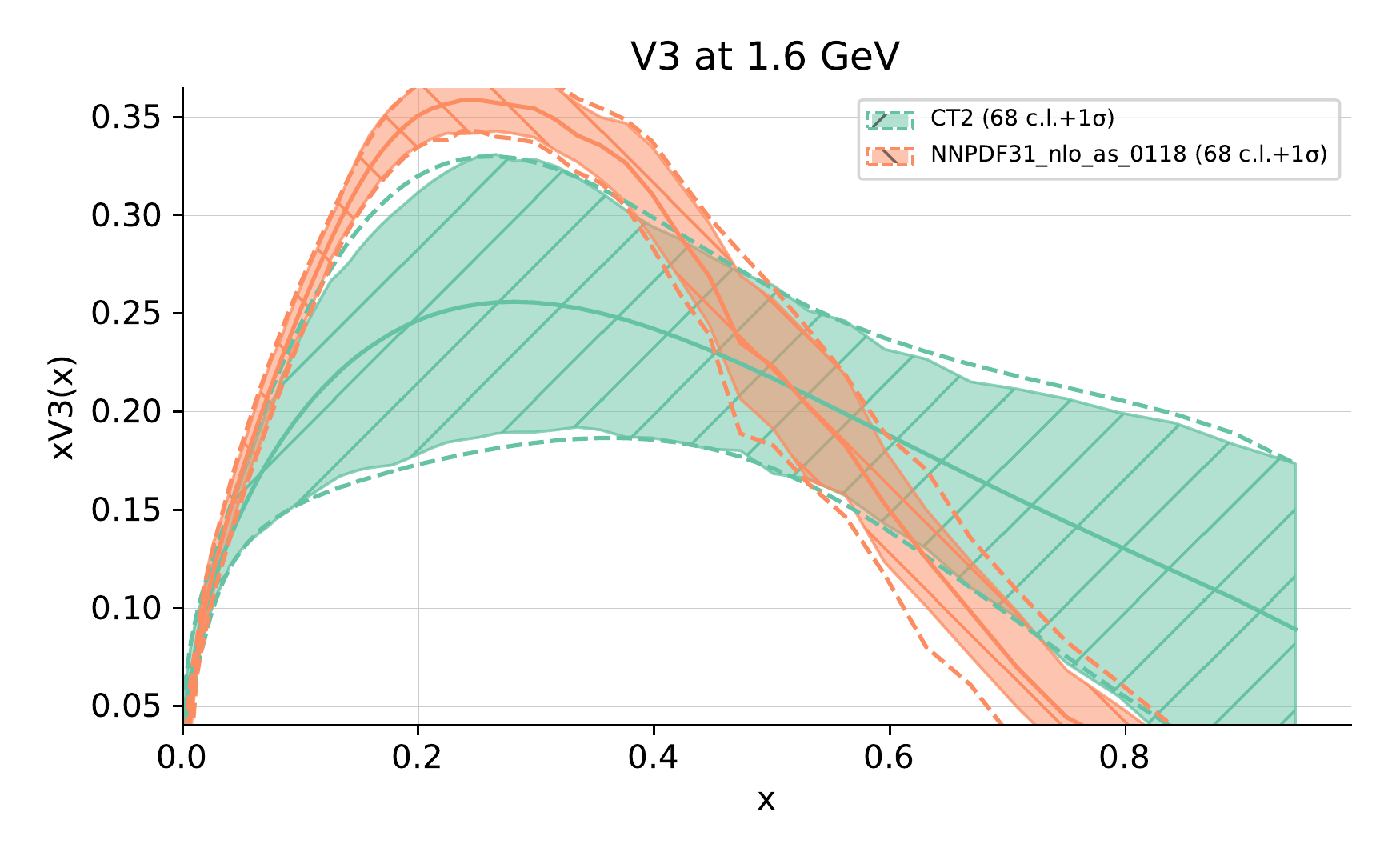}  
	\endminipage\hfill \\
	\minipage{0.45\textwidth} 
	\includegraphics[width=12cm,height=5cm,keepaspectratio]{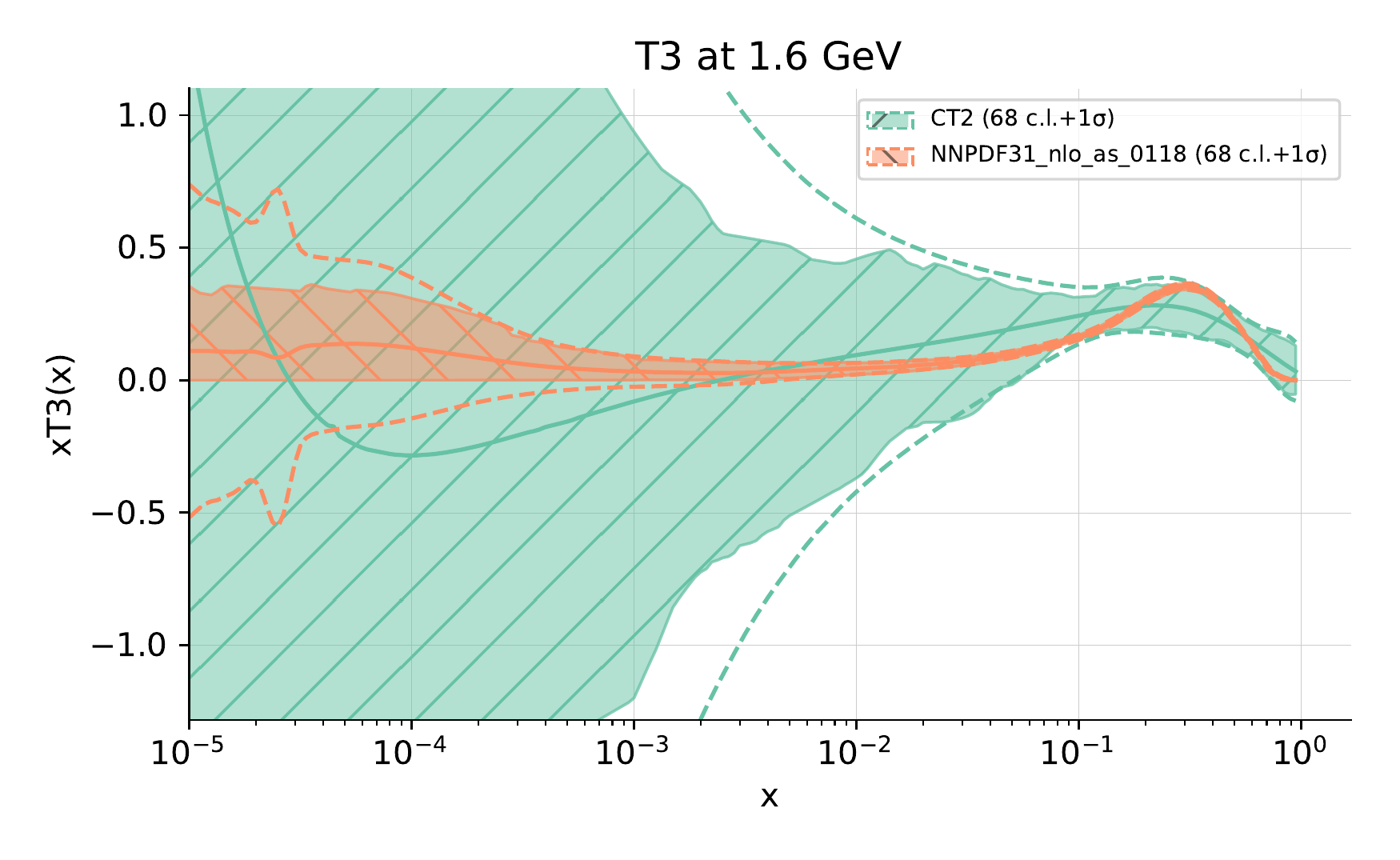}  
	\endminipage\hfill 
	\minipage{0.45\textwidth}
	\includegraphics[width=12cm,height=5cm,keepaspectratio]{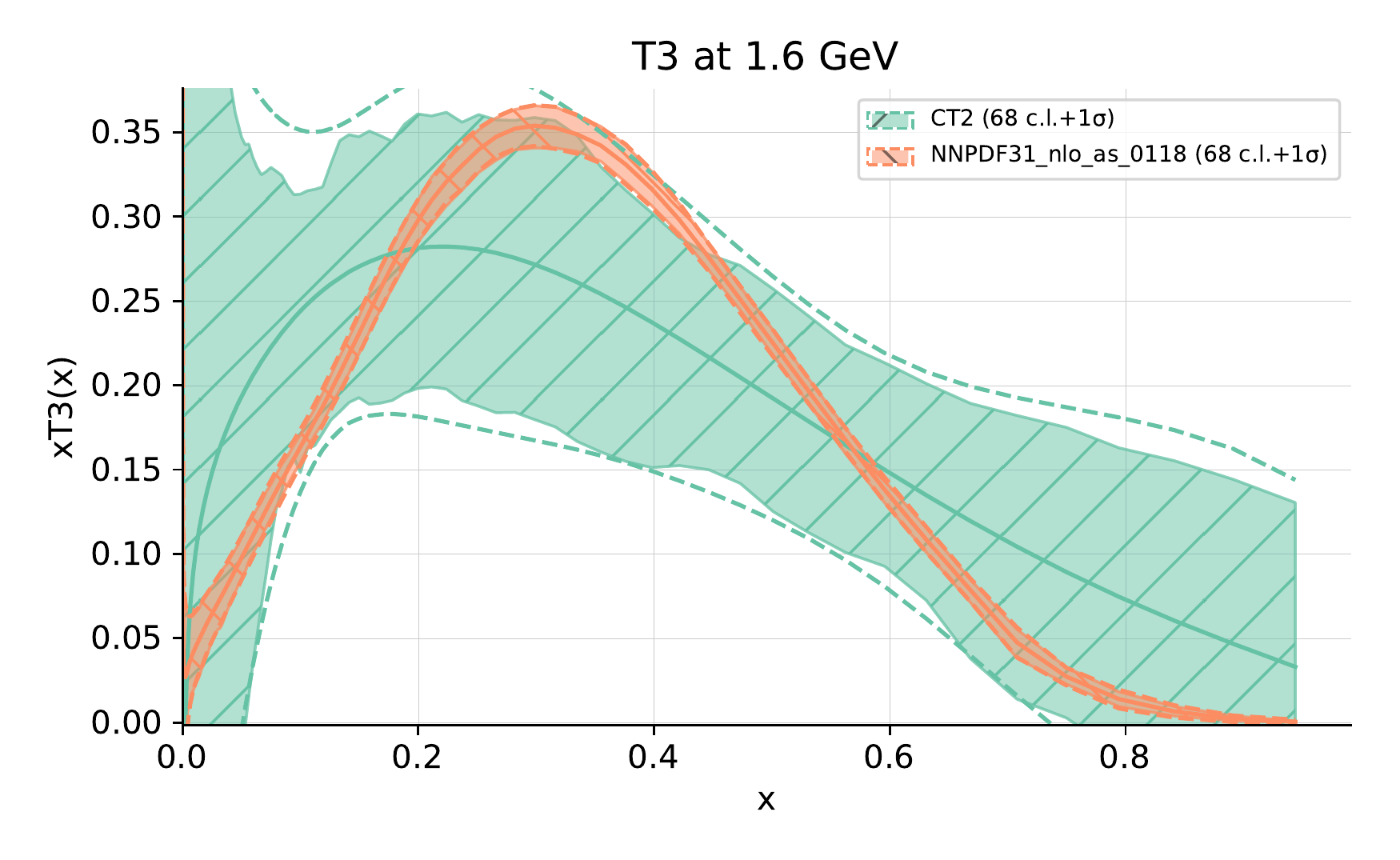}  
	\endminipage\hfill
	\vspace*{-5mm}
	\caption{Closure test fit with real statistical error and no systematics (CT2) compared to the input PDFs set. 
		Top line: $V_3$ combination in linear and logarithmic scale.
		$V_3$ (top line) and $T_3$ (lower line) combinations in linear and logarithmic scale are shown.
		The error band of the reconstructed set is way bigger than the one of the input PDFs,
	showing a non negligible impact of the current statistics over the final PDFs error.}
	\label{fig:2}
\end{figure}

\begin{figure}[h!]
	\minipage{0.45\textwidth}
	\includegraphics[width=12cm,height=5cm,keepaspectratio]{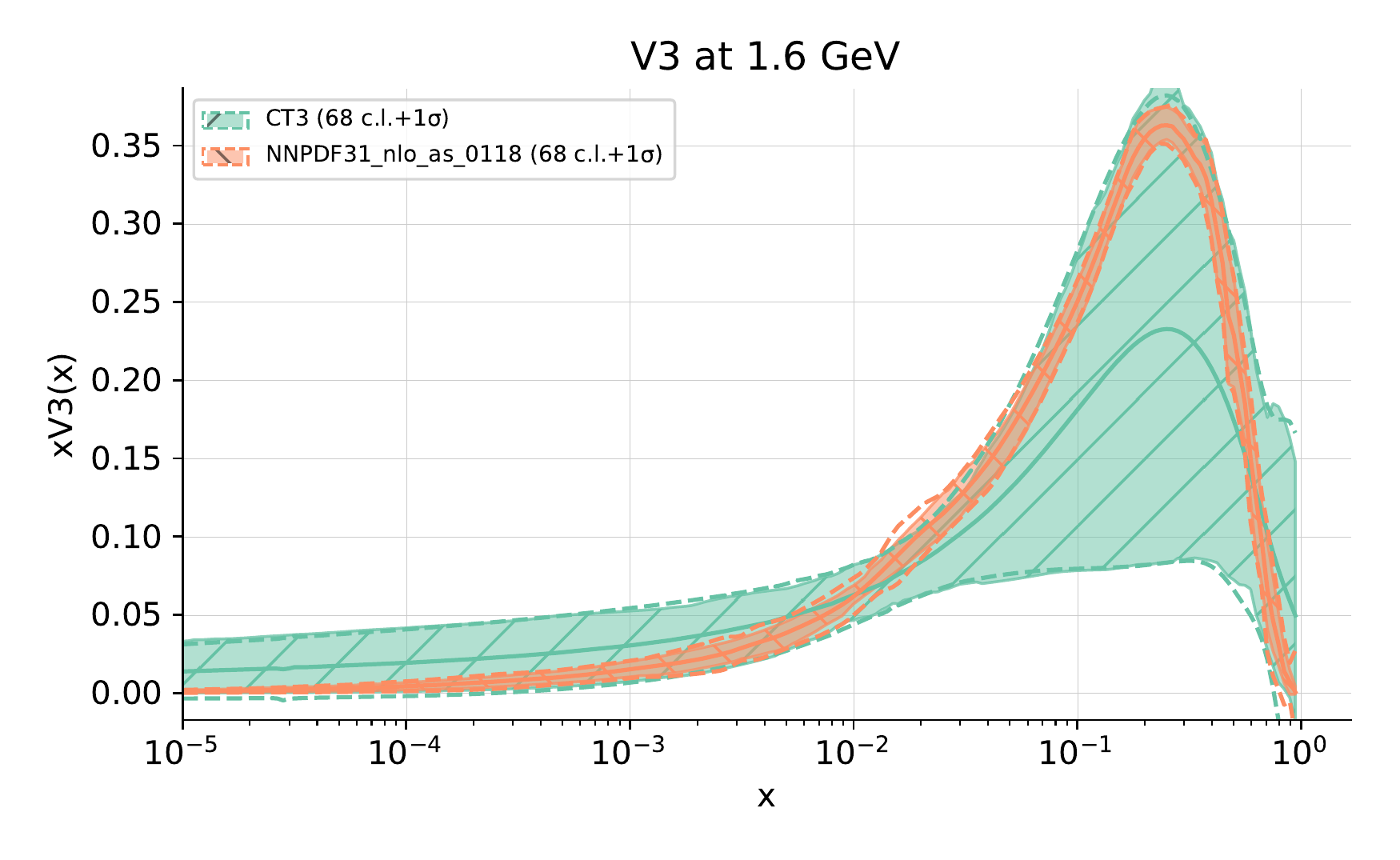}  
	\endminipage\hfill 
	\minipage{0.45\textwidth}
	\includegraphics[width=12cm,height=5cm,keepaspectratio]{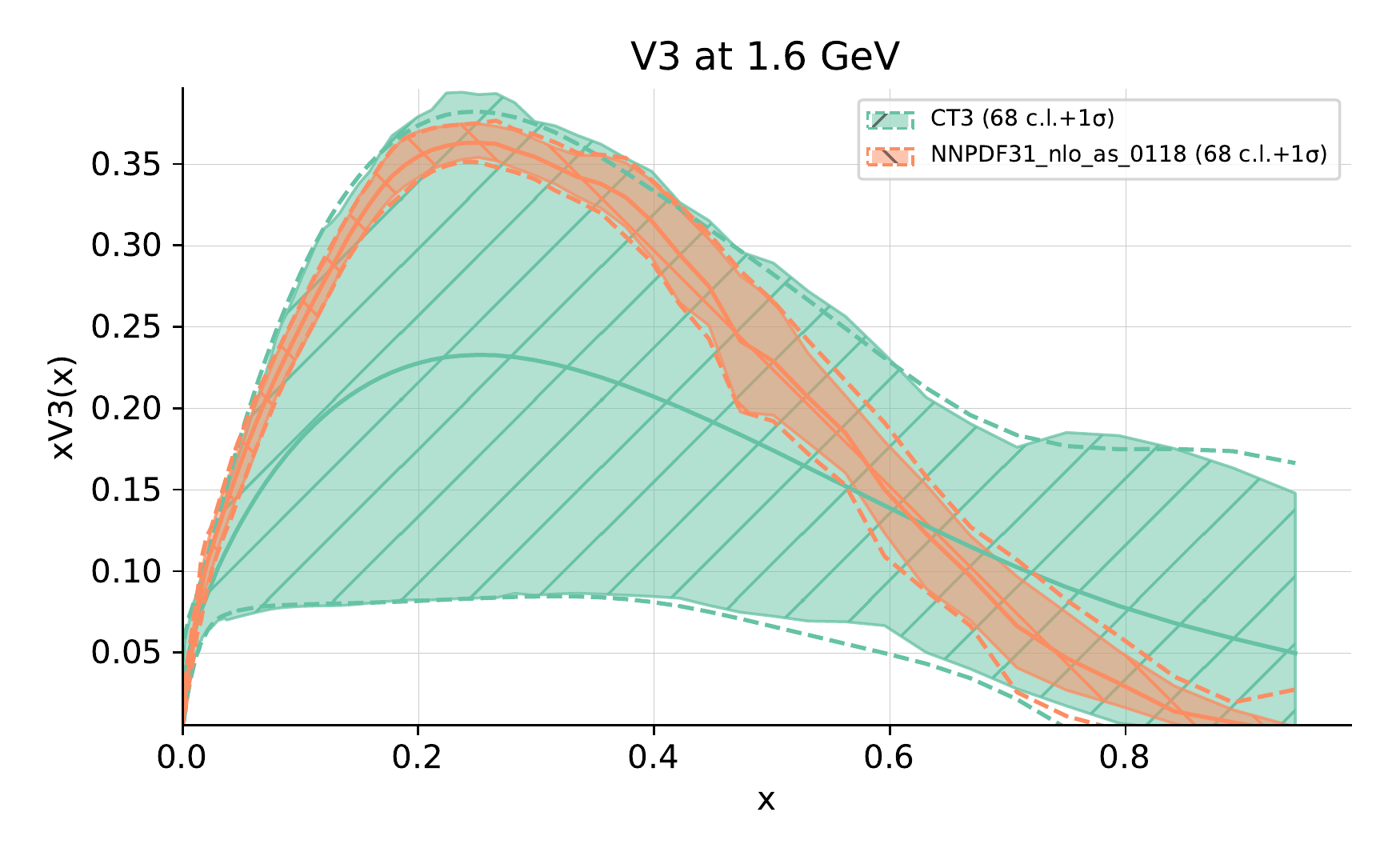}  
	\endminipage\hfill \\
	\minipage{0.45\textwidth} 
	\includegraphics[width=12cm,height=5cm,keepaspectratio]{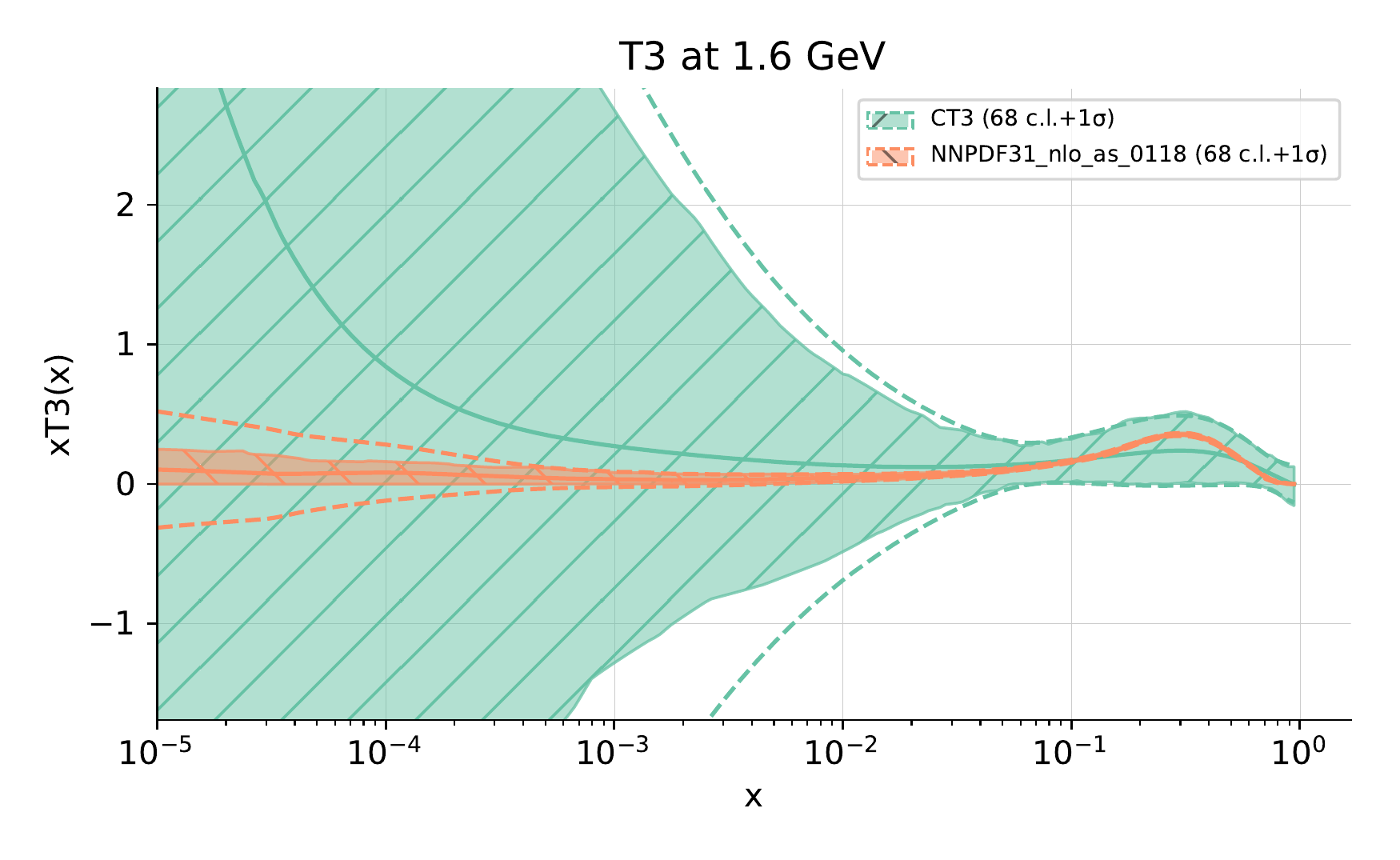}  
	\endminipage\hfill 
        \minipage{0.45\textwidth}
	\includegraphics[width=12cm,height=5cm,keepaspectratio]{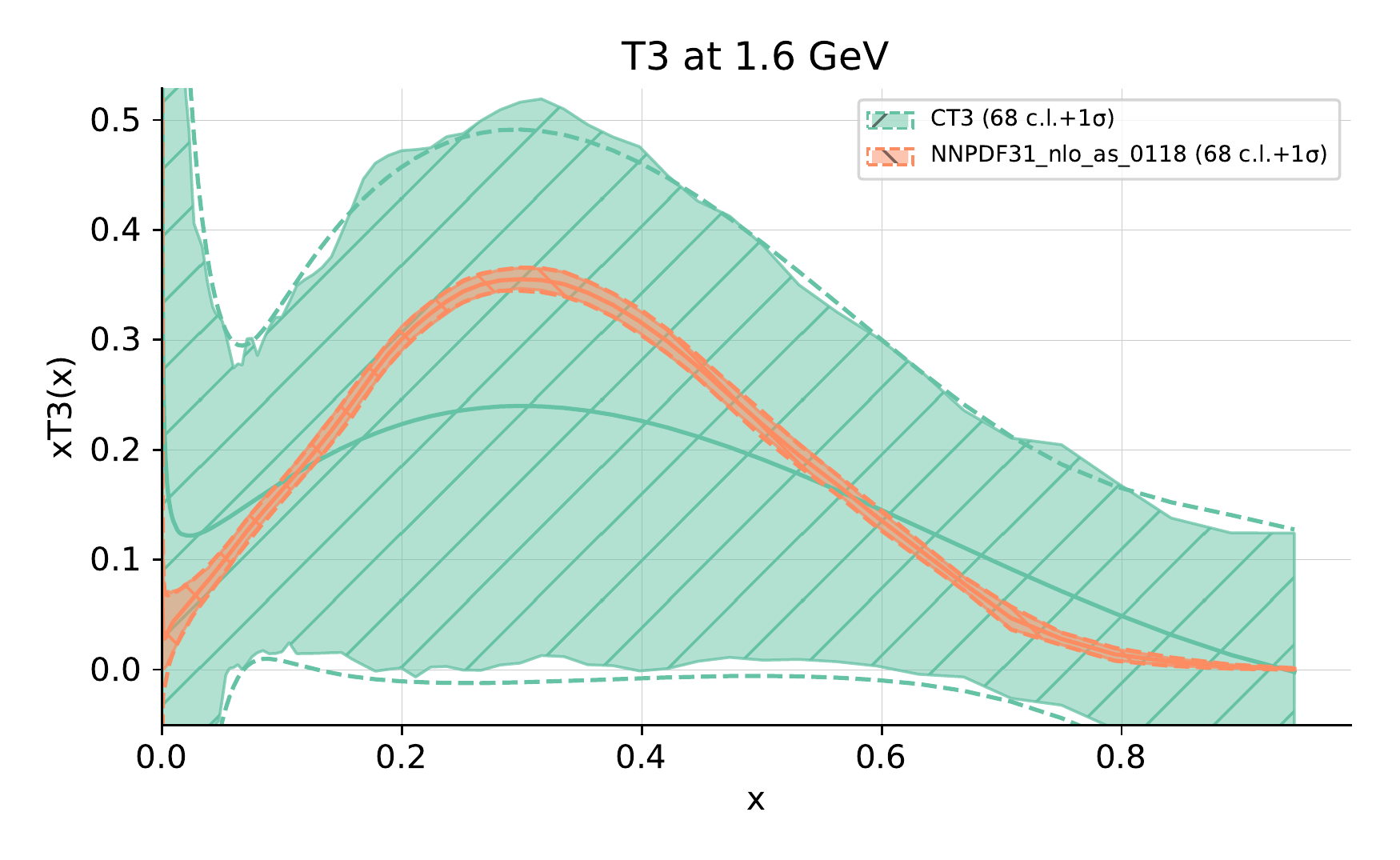}  
	\endminipage\hfill
        \vspace*{-5mm}
	\caption{Closure test fit with real statistical and systematic error (CT3) compared to the input PDFs set. 
                 $V_3$ (top line) and $T_3$ (lower line) combinations in linear and logarithmic scale are shown.
                 The errors of the reconstructed PDFs are huge.}
	\label{fig:3}
\end{figure}

In order to get an idea of the impact on the fit of the statistical and
systematic ME errors, we consider three different scenarios: first we generate
fake data assuming no systematic uncertainties and a small uncorrelated
statistical uncertainty for each point, constant for all of them and of the order of the
smallest real one. From the results of this closure test we can estimate the
real constraining power of the convolution $\circledast$, assuming an ideal
scenario where all the systematics are under control and the statistical error
is kept small. Second, we repeat the exercise but using the real statistical
uncertainties, to assess how much the real statistics of the current simulations
affect the conclusions of the previous case. Finally we look at the effect of
the systematics, considering as a specific example the scenario S2 of
Table~\ref{tab:systematics}. The three cases are summarized in
Table~\ref{tab:CT} and the results are shown in
Figs.~\ref{fig:1},~\ref{fig:2},~\ref{fig:3}.
\begin{table}[h]
\begin{center}
\begin{tabular}{|c|c|c|c|c|}
\hline
Closure test & Statistics & Systematics \\
\hline
{\tt CT1} & fake & - \\
{\tt CT2} & real & - \\
{\tt CT3}  & real  & S2\\
\hline
\end{tabular}
\end{center}
\caption{\label{tab:CT} Closure tests with different choices of the statistical and systematic error. The results for each option above is summarised in the plots below.}
\end{table}

\noindent
Looking at the results for CT1, Fig.~\ref{fig:1}, it is worth stressing that the
lattice data entering the fit are just 16 for the real part and 15 for the
imaginary part of the matrix element. Just half of them are actually used in the
training procedure, while the other ones are used to build the validation set.
In a standard NLO QCD global fit, like the one used as input PDF here, the
number of points entering the analysis is $\mathcal{O}\left(4000\right)$. 
Fig.~\ref{fig:1} shows how good the convolution $\circledast$
is in constraining the PDFs, assuming an ideal scenario where all the
systematics are under control, and the statistics are kept small. Looking at the
results for CT2 and CT3 in Figs.~\ref{fig:2} and \ref{fig:3}, it is clear how
big the impact of the statistical and systematic uncertainties of the ME is on
the PDFs error: in both cases the input PDFs set is reconstructed within 1-sigma
level, with some tension for $V_3$ at medium x in the first case. The PDFs error
is increasingly big, becoming huge when the full systematics are considered.
Fig.~\ref{fig:3} shows what we may expect in a real life scenario.

To sum up, the results from CT1 show how promising this kind of lattice data
might be in constraining PDFs. On the other hand, the results from CT2 and CT3
highlight the importance of having a good control over both the statistical and
systematical uncertainties in the lattice simulations of the ME. It is worth
noticing, however, that the overall error band of the reconstructed PDFs, even in
presence of the full systematic errors, would surely be reduced when new data
are available.

\subsection{Fit results}
\label{subsec:fits}
In this section, we present our results for fits ran over the data from
Refs.~\cite{Alexandrou:2018pbm,Alexandrou:2019lfo}, described in Sec.~\ref{subsec:latticedata}. As
mentioned before, we consider 6 different scenarios for the treatment of the
systematic errors, summarized in Table~\ref{tab:systematics}. We show results
for "optimistic" (S1,S4), "realistic" (S2,S5) and "pessimistic" (S3,S6)
scenarios, the difference between the elements of each couple being the nature
of the systematic errors: an additive shift given by a percentage of the ME for
the first, a constant shift for all the ME points for the second one. 

\begin{figure}[h]
	\minipage{0.45\textwidth}
	\includegraphics[width=12cm,height=5cm,keepaspectratio]{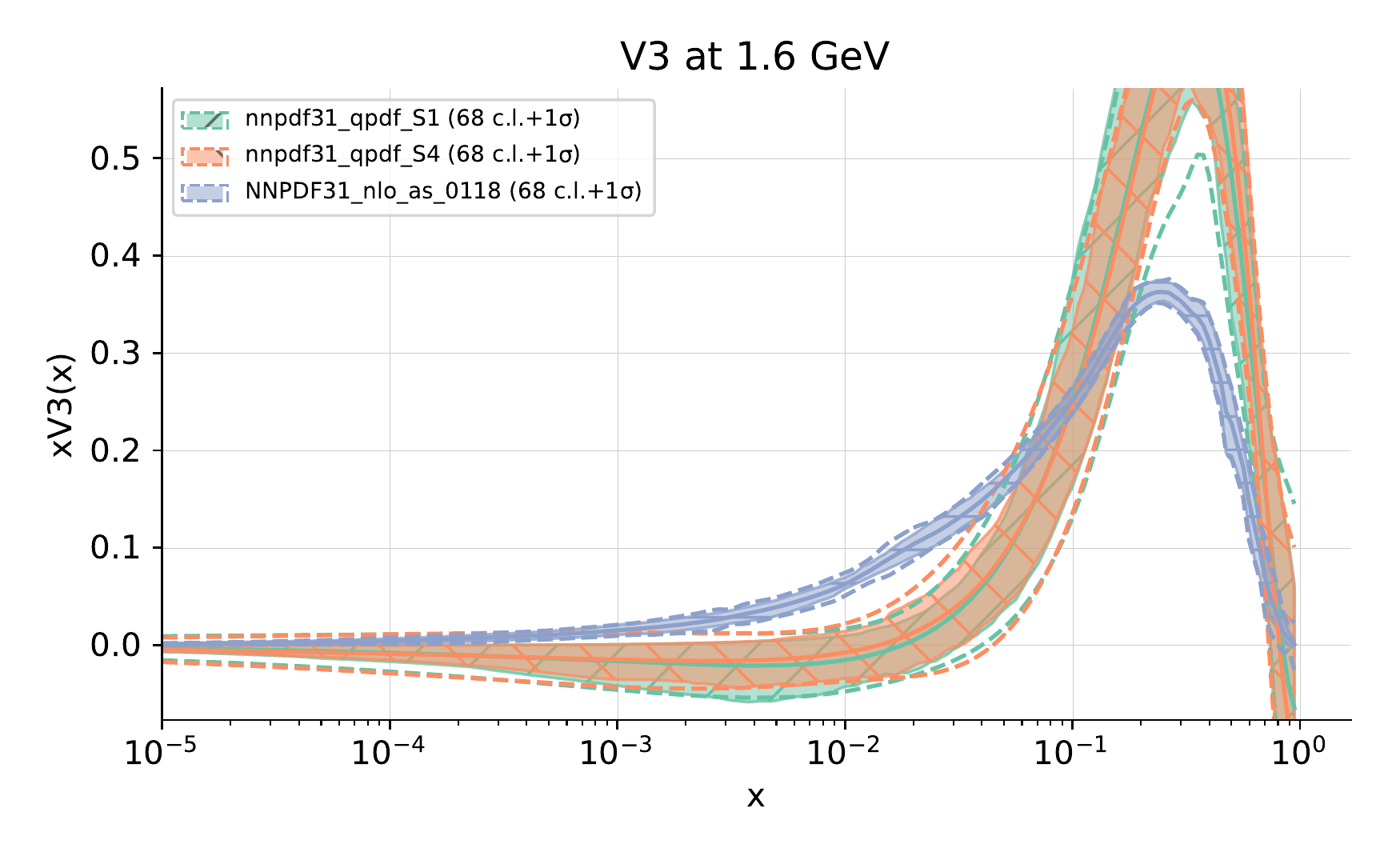}  
	\endminipage\hfill 
	\minipage{0.45\textwidth}
	\includegraphics[width=12cm,height=5cm,keepaspectratio]{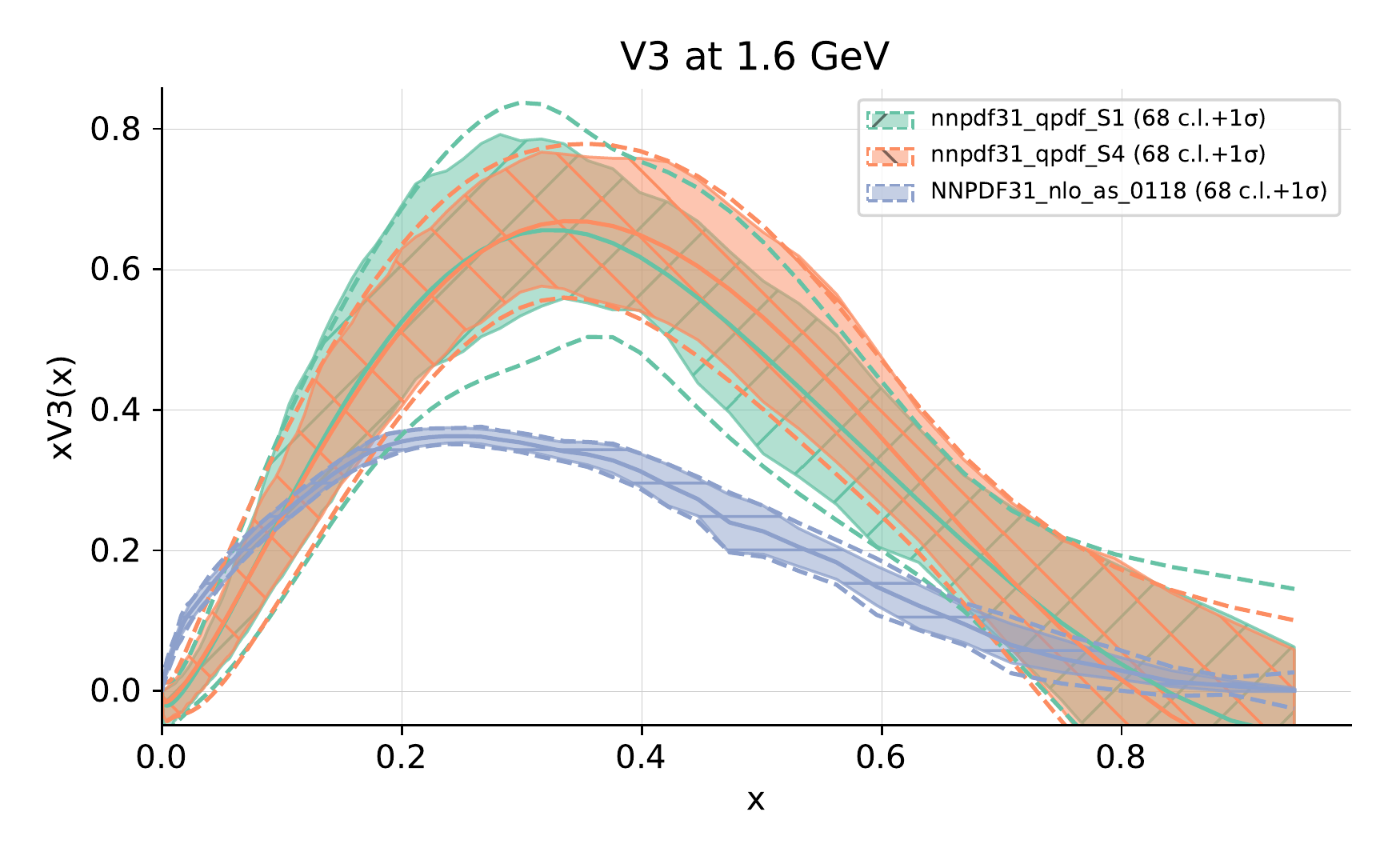}  
	\endminipage\hfill \\
	\minipage{0.45\textwidth}
	\includegraphics[width=12cm,height=5cm,keepaspectratio]{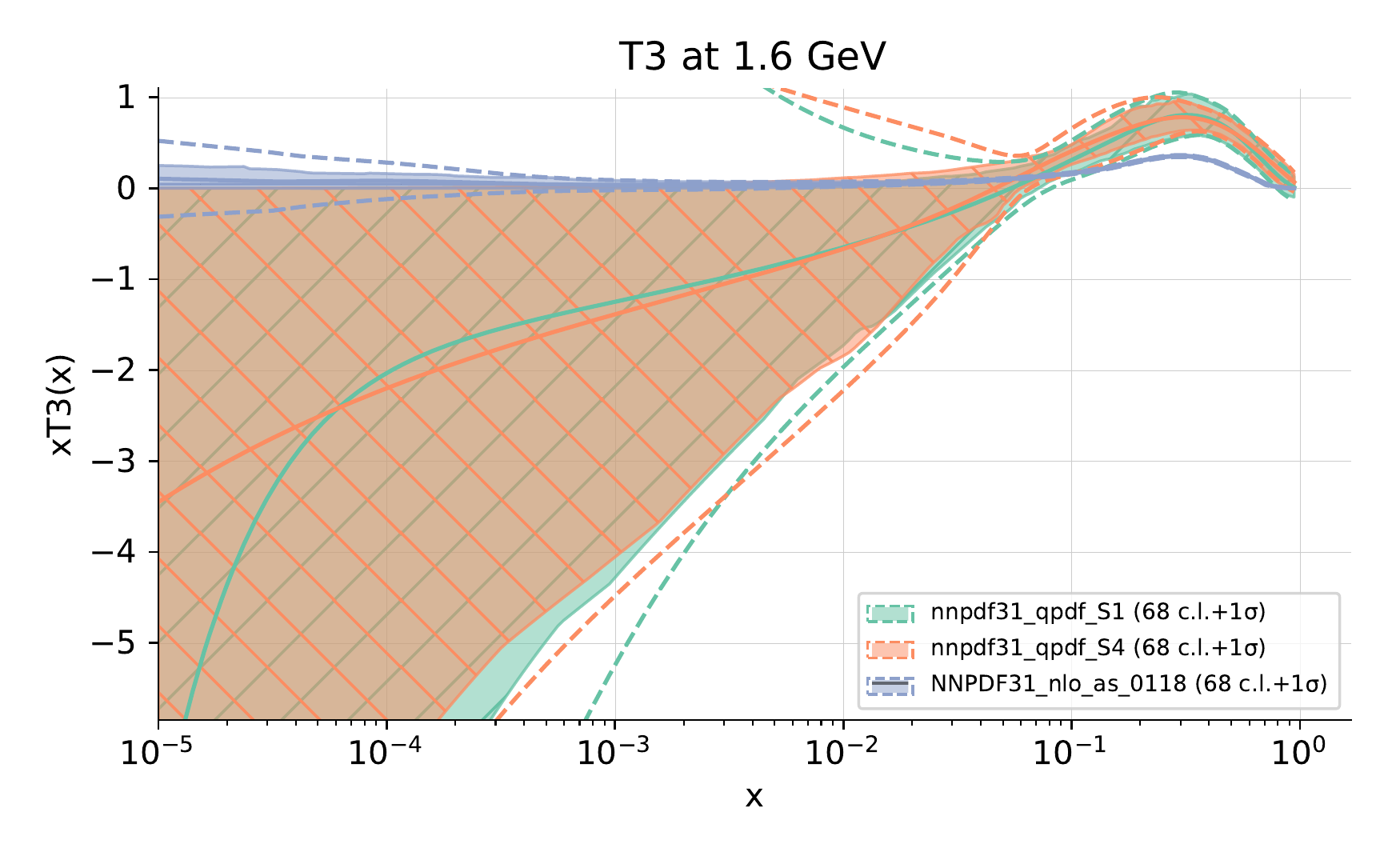}  
	\endminipage\hfill 
	\minipage{0.45\textwidth}
	\includegraphics[width=12cm,height=5cm,keepaspectratio]{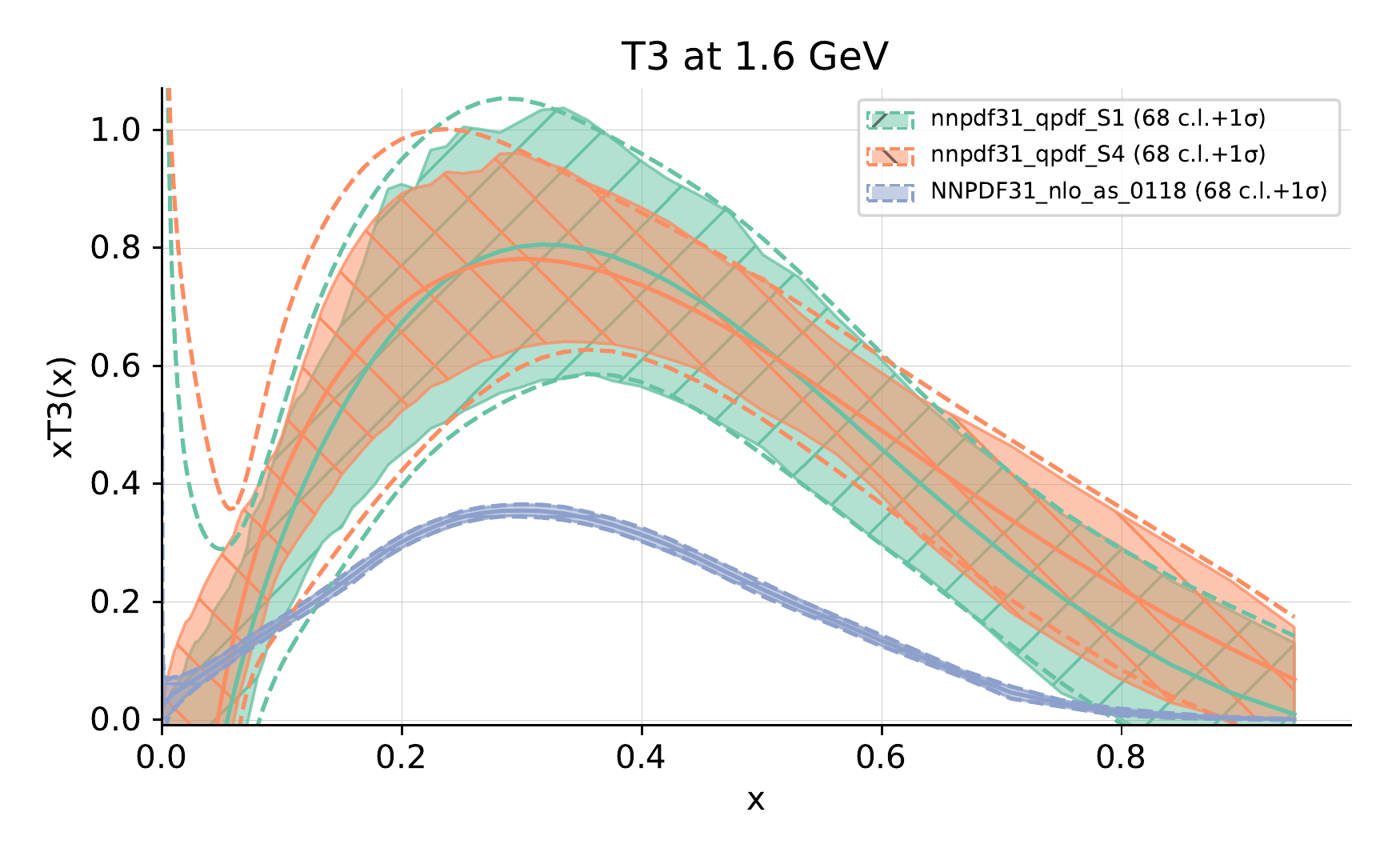}  
	\endminipage\hfill \\
	\vspace*{-5mm}
	\caption{S1 vs.\ S4: S1 results are sligthly more conservative than the S4
	ones, but overall there is no significant difference between the two optimistic
	scenarios.}
	\label{fig:4}
\end{figure}

\begin{figure}[h!]
	\minipage{0.45\textwidth}
	\includegraphics[width=12cm,height=5cm,keepaspectratio]{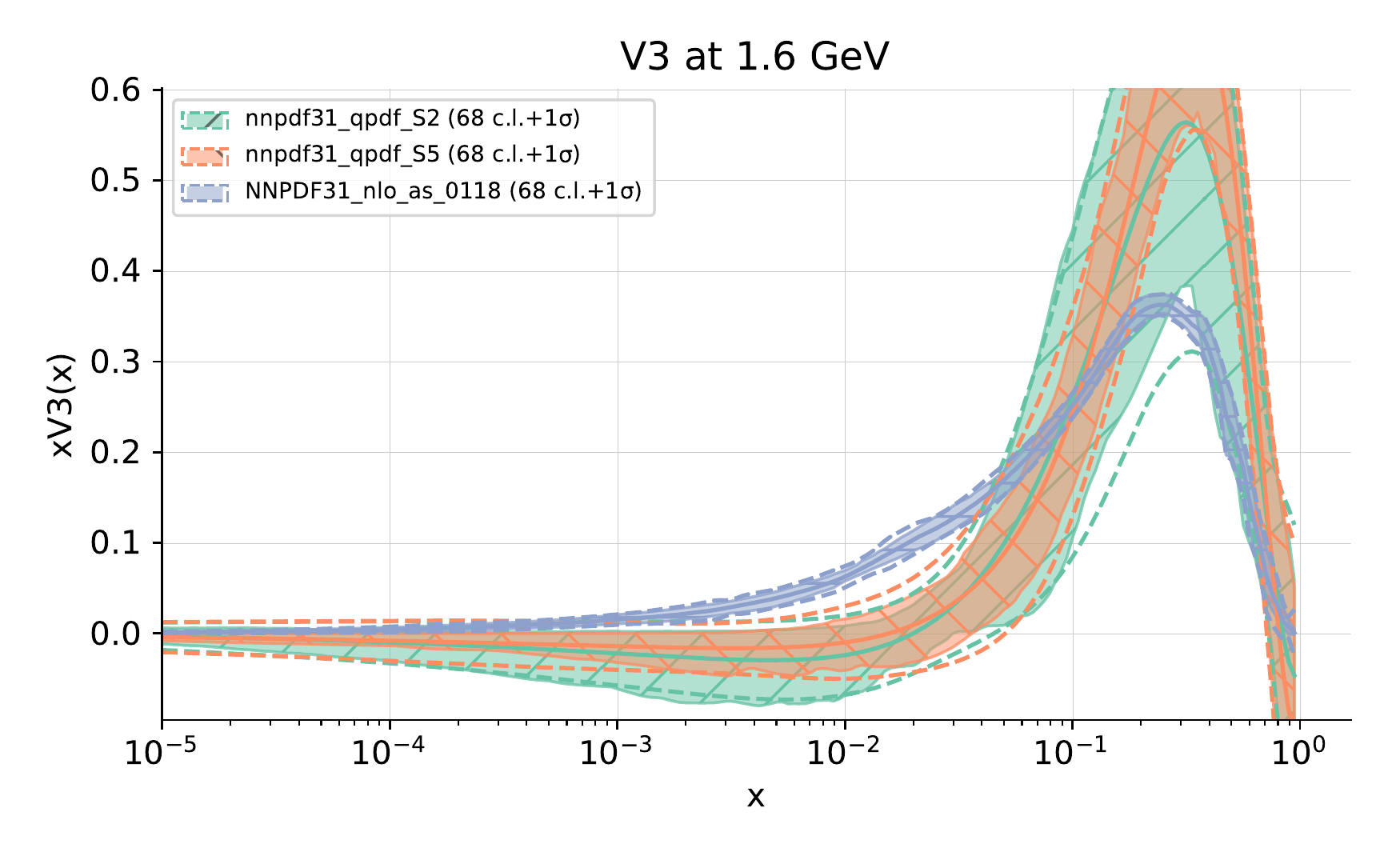}  
	\endminipage\hfill 
	\minipage{0.45\textwidth}
	\includegraphics[width=12cm,height=5cm,keepaspectratio]{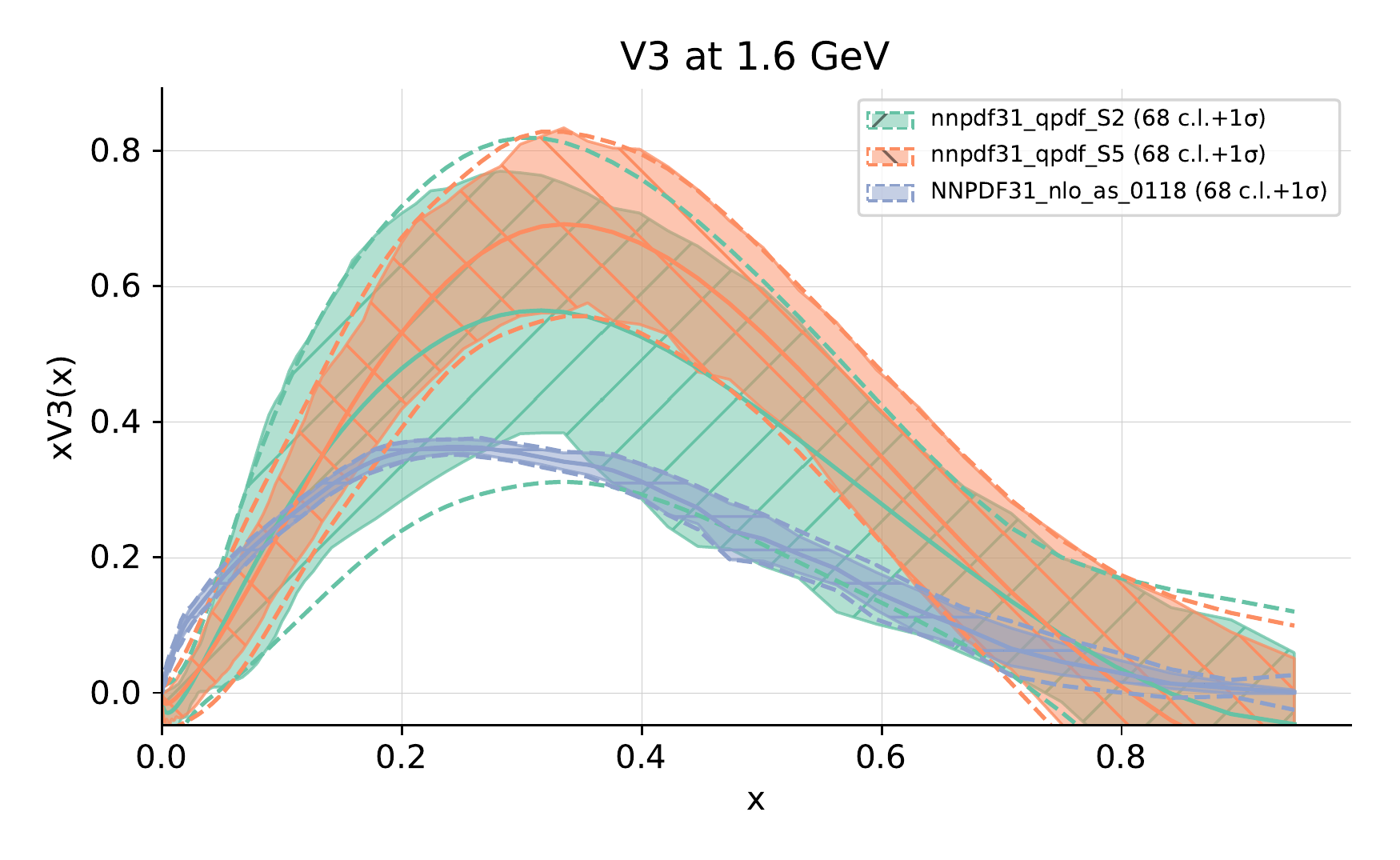}  
	\endminipage\hfill \\
        \minipage{0.45\textwidth}
	\includegraphics[width=12cm,height=5cm,keepaspectratio]{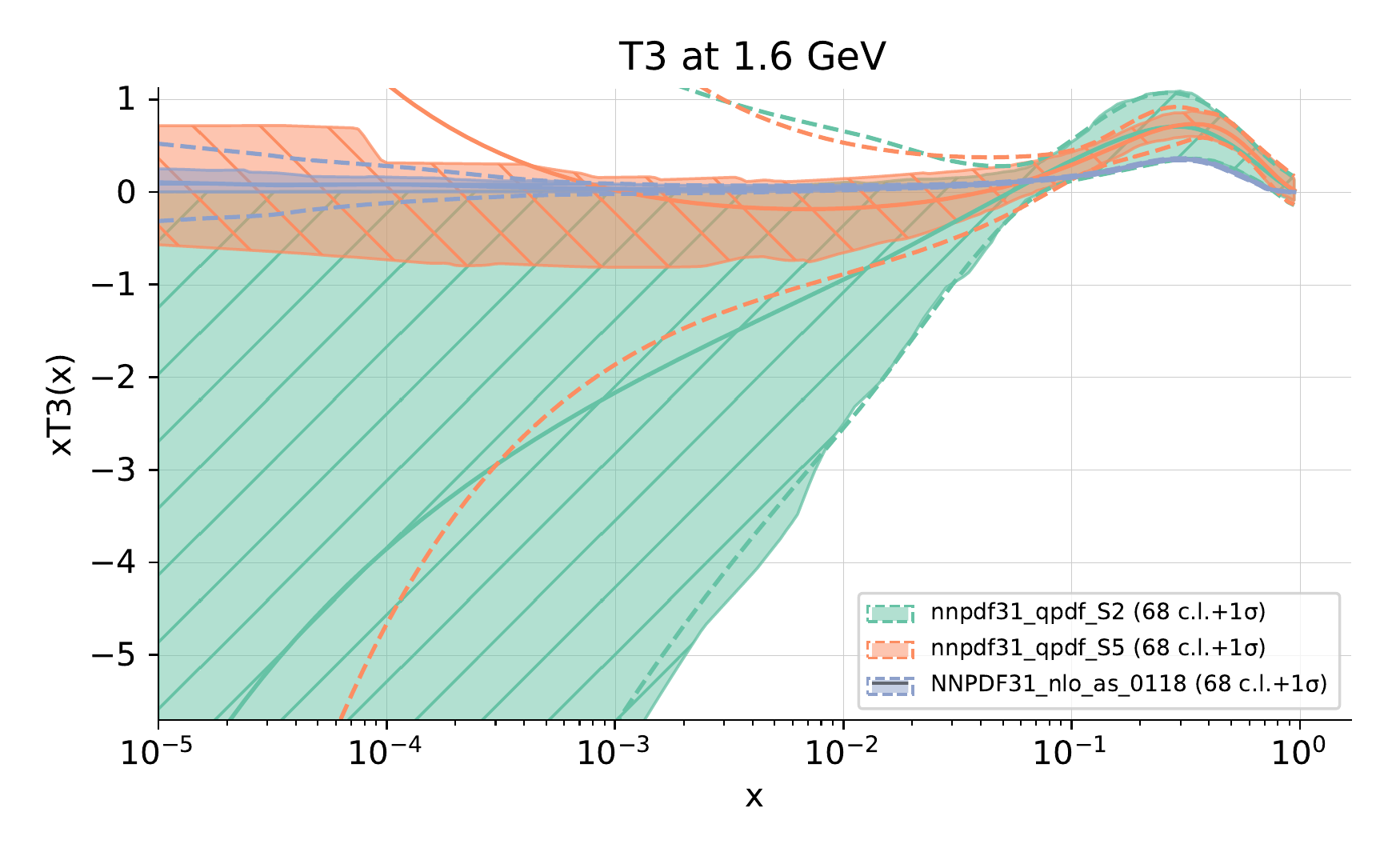}  
	\endminipage\hfill 
	\minipage{0.45\textwidth}
	\includegraphics[width=12cm,height=5cm,keepaspectratio]{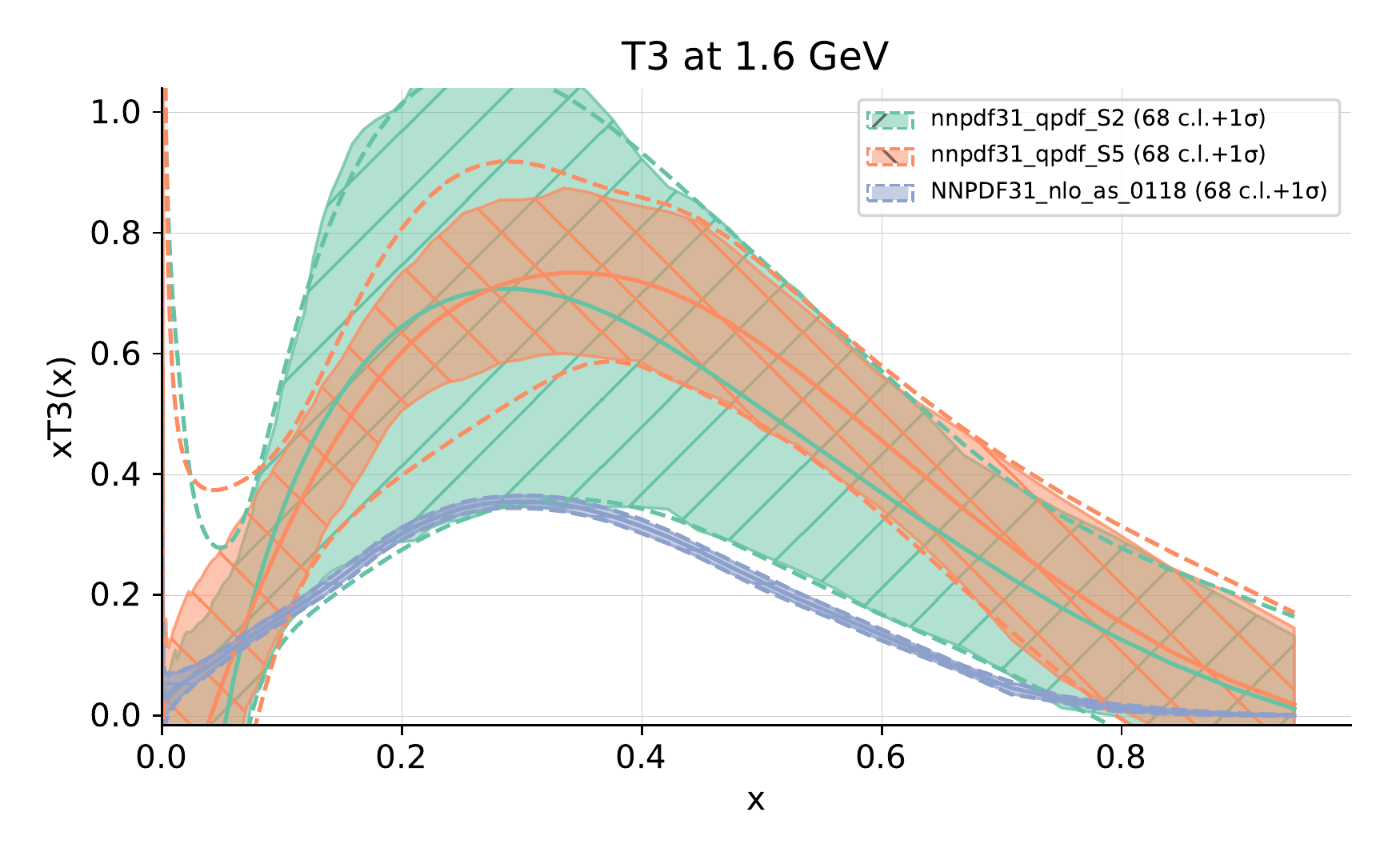}  
	\endminipage\hfill \\
        \vspace*{-5mm}
	\caption{S2 vs.\ S5: S2 results are more conservative than the S5 ones,
        showing also a small shift of the replica 0 towards the light-cone PDFs.
        Overall, S2 results are comptible with {\tt NNPDF31\_nlo\_0118} within
        1-sigma level.}
	\label{fig:5}
\end{figure}

\begin{figure}[h!]
	\minipage{0.45\textwidth}
	\includegraphics[width=12cm,height=5cm,keepaspectratio]{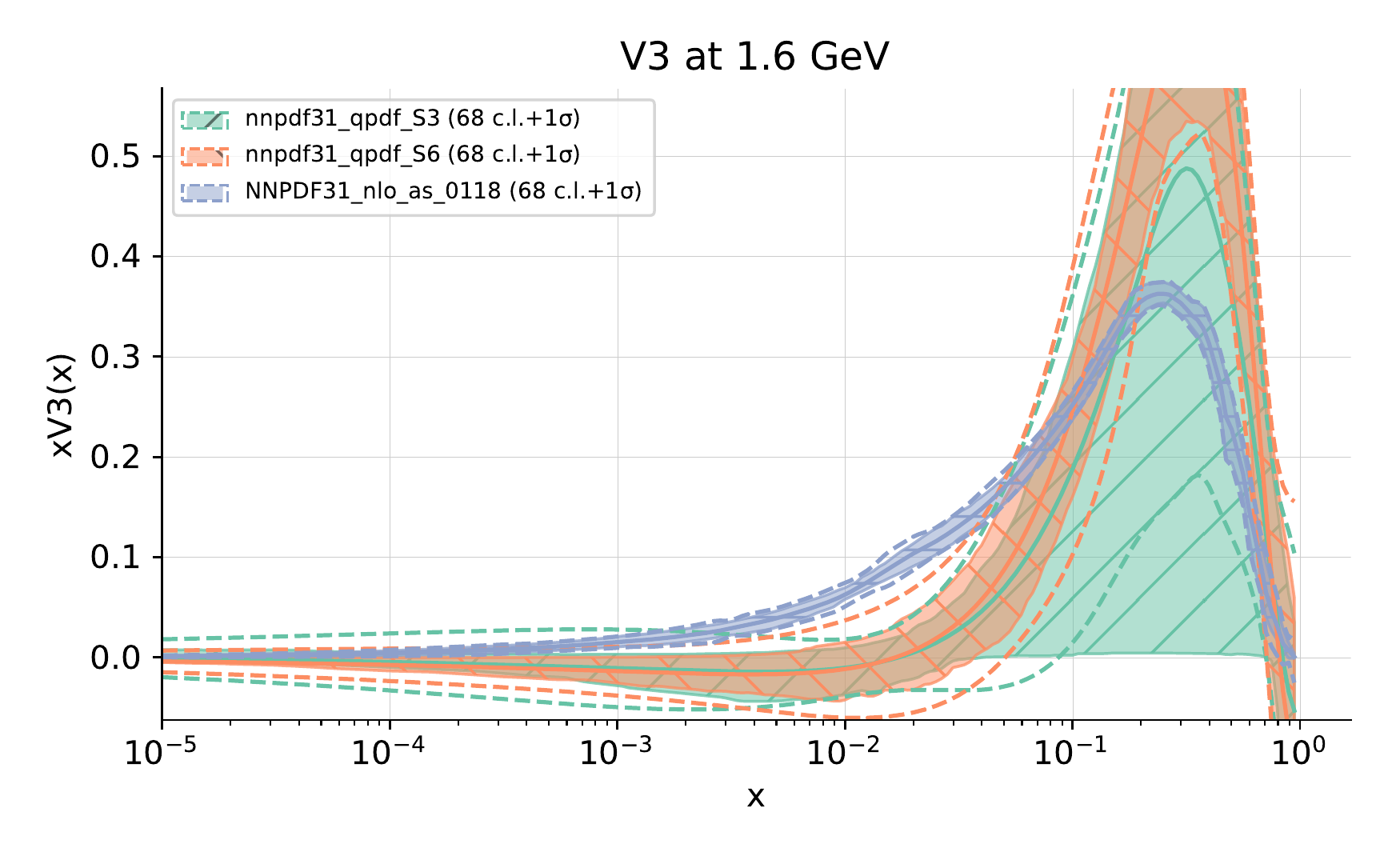}  
	\endminipage\hfill 
	\minipage{0.45\textwidth}
	\includegraphics[width=12cm,height=5cm,keepaspectratio]{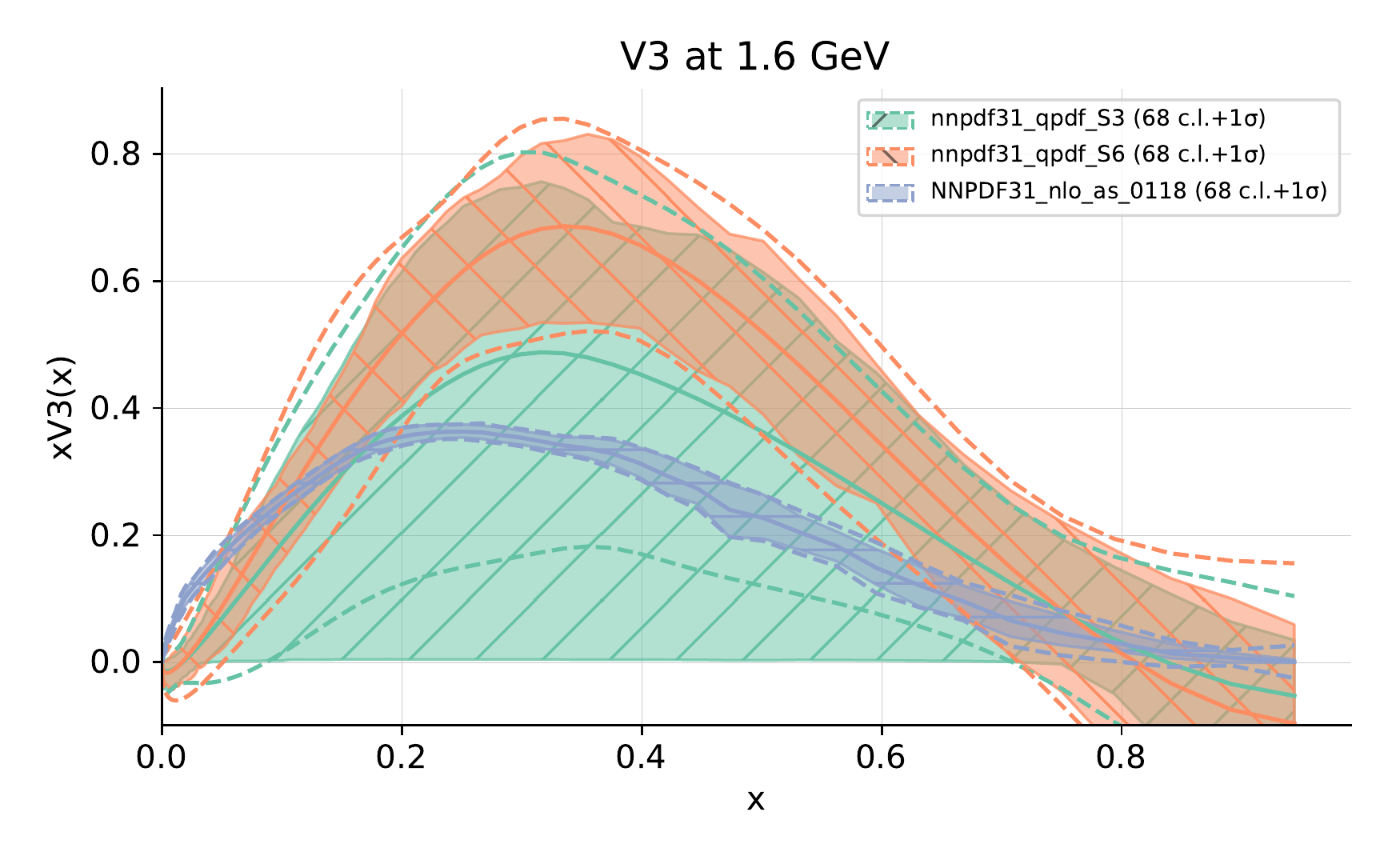}  
	\endminipage\hfill \\
        \minipage{0.45\textwidth}
	\includegraphics[width=12cm,height=5cm,keepaspectratio]{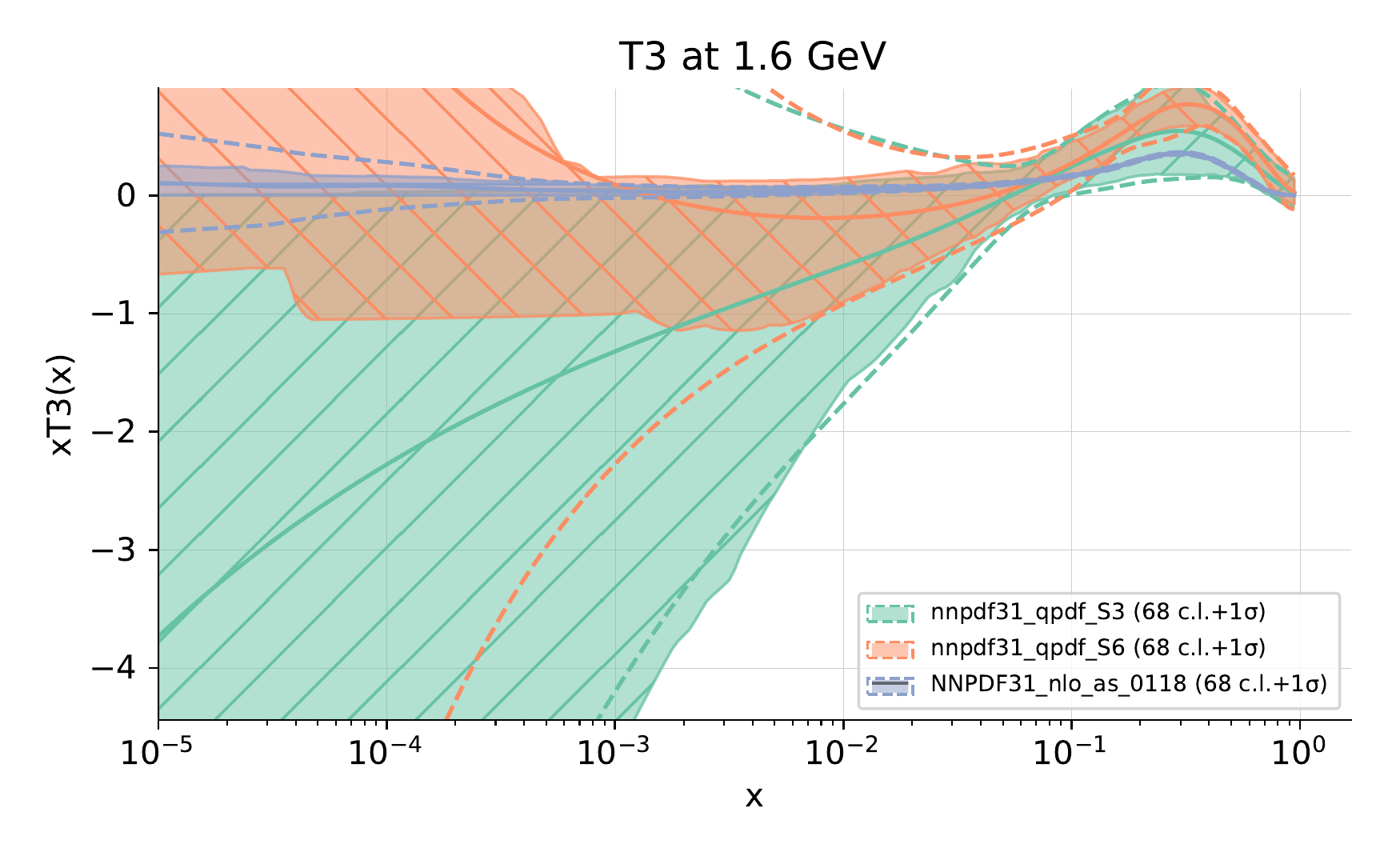}  
	\endminipage\hfill 
	\minipage{0.45\textwidth}
	\includegraphics[width=12cm,height=5cm,keepaspectratio]{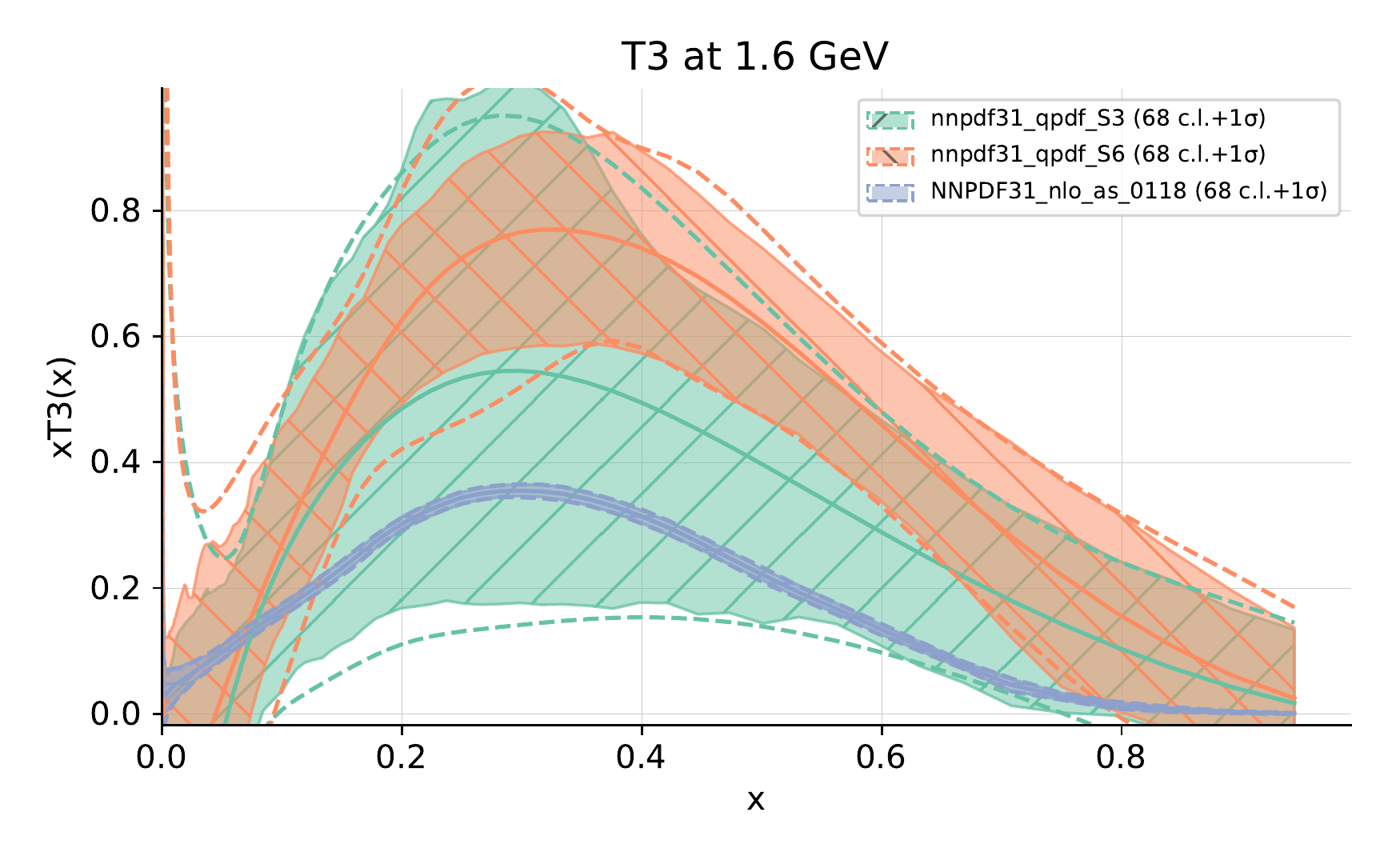}  
	\endminipage\hfill \\
        \vspace*{-5mm}
	\caption{S3 vs.\ S6: S3 results are extremely conservative, while those for S6
        do not show a qualitative difference with respect to S4 and S5.}
	\label{fig:6}
\end{figure}

The results of the fit for the two optimistic scenarios are shown in
Fig.~\ref{fig:4}. S1 is sligthly more conservative than S4, but overall there is
not much difference between them. The situation changes for the more realistic
scenarios (Fig.~\ref{fig:5}), where S2 is much more conservative than S5. In the
former case the tension with {\tt NNPDF31\_nlo\_0118} is smaller than what we
observe in the previous scenarios, due to the increase in the error band and to
a slight shift of the central replica of the fit. Similar comments can be made
for the most pessimistic scenarios, shown in Fig.~\ref{fig:6}, having S3 with a
huge error band and a more remarkable shift of the central replica towards the
one of NNPDF31. Overall, we notice how, when the systematics are given by a
percentage of the ME, we get qualitatively different results moving from one
scenario to the other, while in the case we consider constant shifts there is no
much difference between different cases. 
 
\clearpage
\newpage

\section{Conclusions and future work}
\label{sec:summary}
In this work, we have used the factorization of quasi-PDFs in order to relate
the unpolarized isovector parton distribution to well-defined matrix elements
computable on the lattice. Using some of the currently available lattice data,
we have used such result to extract the nonsinglet distributions $V_3$ and $T_3$
within the {\tt NNPDF} framework, studying also different possible scenarios for
the treatment of the systematic uncertainties from lattice QCD simulations.

Our first results from closure tests show how effective these lattice data might
be in constraining PDFs, allowing a consistent determination of the target
distribution starting from $\mathcal{O}\left(15\right)$ ME points. On the other
hand, we show that a consistent treatment of the lattice systematics is
extremely important, and how the final result of the fit strongly depends on the
specific systematics scenario we consider. Considering the most realistic ones,
agreement with the phenomenological PDFs is observed within 1 sigma level, for
both the nonsiglet distributions considered here. The error bands are, however,
very large with respect to the corresponding phenomenological PDFs, showing again how
important the control over the lattice systematics is. 

Despite having focused on the quasi-PDFs case, the framework we
implemented is general enough to allow the inclusion in the same analysis of
data coming from different lattice simulations. One direction for future work
might be a similar analysis repeated considering, for example, pseudo-PDFs data
rather than quasi-PDFs ones, to see weather the conclusions we got here apply
also to data coming from different lattice approaches. A fully global lattice
QCD fit, which would be easily implemented within this framework, would then be
the next logical step: it would then be interesting to see how much the
inclusion of more lattice data in the same analysis affects the error band of
the resulting PDFs, and whether or not a better agreement with the PDFs
determined from experimental data is reached.

Our approach allows also to combine in the same analysis both experimental and
lattice data. In principle we could then study the impact of the available
lattice data on an existing PDF set, like {\tt NNPDF31}. This, however, would
require a deeper control and assessment of the lattice systematics involved in
the simulations, together with a suitable way of treating the theoretical errors
involved in the computation of the theory predictions entering the fits (missing
higher orders in the matching coefficents and power correction terms), to avoid
biasing the global QCD fit. This will be done in a future work.

Finally, we note how the current analysis can be extended without any additional
complications to the other nonsinglet distributions defined in
Eq.~\eqref{eq:fADef}, as soon as lattice data for the corresponding matrix
elements become available. 

\subsection*{Acknowledgments}
Our thinking on this subject has been sharpened by discussions with K~Orginos
and C~Monahan. We are thankful to V~Bertone, S~Carrazza, ER~Nocera, J~Rojo, G~Bali and V~Braun
for useful discussions and technical help. We thank the Extended Twisted Mass
Collaboration for sharing the lattice data that were originally produced for
Refs.~\cite{Alexandrou:2018pbm,Alexandrou:2019lfo}. Last but not least, we are
grateful to the organisers of 'The Second Workshop on Parton Distribution
Functions' in Taiwan in November 2018. This work is the result of discussions at
that meeting. 

TG is supported by The Scottish Funding Council, grant H14027. KC is supported
by National Science Centre (Poland) grant SONATA BIS 2016/22/E/ST2/00013. LDD is
supported by an STFC Consolidated Grant, ST/P0000630/1, and a Royal Society
Wolfson Research Merit Award, WM140078.

\appendix

%
\section{Matching coefficient and lattice convolution}
\label{app:coefficients}
As detailed at the end of Sec.~\ref{sec:theoryoverview}, the matching coefficients to be
used to relate the data of Refs.~\cite{Alexandrou:2018pbm,Alexandrou:2019lfo} to the light-cone PDFs
are those expressed in the $\MMSb$ scheme. Their explicit expression is given
by \cite{Alexandrou:2018pbm,Alexandrou:2019lfo}
\begin{equation}
	\label{eq::matching}
	\begin{split}
		&C_{3}\left(\xi,\eta \right)= \delta\left(1-\xi\right)\, +  C_{3}^{\text{NLO}}\left(\xi,\eta \right)_+\,,  \\ \\ 
		& C_{3}^{\text{NLO}}\left(\xi,\eta \right)_+ =  \frac{\alpha_s}{2\pi}C_F \begin{cases} \left[\frac{1+\xi^2}{1-\xi}\log{\frac{\xi}{\xi-1}} + 1 + \frac{3}{2\xi}\right]_{+(1)} \,\,\,\,\,\,\,\,\,\,\,\,\,\,\,\,\,\,\,\,\,\,\,\,\,\,\,\,\,\,\,\,\,\,\,\,\,\,\,\,\,\,\,\,\,\,\,\,\,\,\,\,\,\,\,\,\,\,\,\, \xi > 1\\ \left[\frac{1+\xi^2}{1-\xi}\log\left[{\frac{1}{\eta^2}}\left(4\xi\left(1-\xi\right)\right)\right]  -\frac{\xi\left(1+\xi\right)}{1-\xi}\right]_{+(1)}\,\,\,\,\,\,\,\,\,\,\,\,\,\,\,\,\,\,\, 0<\xi < 1 \\ \left[-\frac{1+\xi^2}{1-\xi}\log{\frac{\xi}{\xi-1}} - 1 + \frac{3}{2\left(1-\xi\right)}\right]_{+(1)} \,\,\,\,\,\,\,\,\,\,\,\,\,\,\,\,\,\,\,\,\,\,\,\,\,\,\,\,\,\,\,\,\,\,\,\,\,\,\,\,\,\,\,\,\,\, \xi<0\end{cases}\, .
	\end{split}
\end{equation}
The matching coefficients relate the light-cone PDF to the quasi-PDF up to power suppressed terms
according to
\begin{align}
	\label{eq::matchingfud}                                   
	\tilde{f}_{3}\lp x P_z, {\mu}^2 \rp =              
	\int_{-1}^{1} \frac{dy}{|y|}\, C_{3}\lp\frac{x}{y},\frac{\mu}{y P_z}\rp  
	f_{3}\lp y, {\mu}^2\rp. 
\end{align}
In the following, we work out the full expression of the coefficients appearing in Eqs.~\eqref{eq::V3factorization}, \eqref{eq::T3factorization}.
Starting from Eq.~\eqref{eq::matchingfud} we have
\begin{align}
	\label{eq::explicitmatching}
	\tilde{f}_3\left(x,\mu^2,P_z\right) = \int_{-1}^{1} \frac{dy}{|y|}\, \delta\lp 1- \frac{x}{y} \rp  
	f_{3}\lp y, {\mu}^2\rp + \int_{-1}^{1} \frac{dy}{|y|}\, C_{3}^{\text{NLO}}\lp\frac{x}{y},\frac{\mu}{y P_z}\rp_+  
	f_{3}\lp y, {\mu}^2\rp\, .
\end{align}
Let us focus on the next-to-leading order term, making the plus distribution
explicit. In order to do so, we find it useful to split the integral in the two
contributions for $y<0$ and $y>0$. A change of variables, $\smallfrac x y =
\xi$, yields 
\begin{align}
	\tilde{f}_3^{\text{NLO}}\left(x,\mu^2,P_z\right) 
	  & \equiv \int_{-1}^{1} \frac{dy}{|y|}\, 
	C_{3}^{\text{NLO}}\lp\frac{x}{y},\frac{\mu}{y P_z}\rp_+ 
	f_{3}\lp y, \mu^2\rp \nonumber \\ 
	  & = \int_{|x|}^{\infty} d\xi\,          
	C_3^{\text{NLO}}\left(\xi,\frac{\mu\xi}{x P_z}\right)_+ 
	\frac{1}{|\xi|} f_3\left(\frac{x}{\xi},\mu^2\right) + \nonumber \\
	  & \quad + \int_{-\infty}^{-|x|} d\xi\,  
	C_3^{\text{NLO}}\left(\xi,\frac{\mu\xi}{xP_z}\right)_+ 
	\frac{1}{|\xi|} f_3\left(\frac{x}{\xi},\mu\right). 
\end{align}
The plus distribution appearing in the matching coefficients is implemented as
follows
\begin{align}
	\int_{x}^{\infty} d\xi\, 
	C\left(\xi,g\left(\xi\right)\right)_+ f\left(\xi\right) 
	= \int_{x}^{\infty} d\xi\,
	\left[C\left(\xi,g\left(\xi\right)\right)f\left(\xi\right) - 
	C\left(\xi,g\left(1\right)\right)f\left(1\right) \right]\, , 
\end{align}
with $g\lp\xi\rp =\smallfrac{\mu\xi}{xP_z} $, so that
\begin{align}
	\tilde{f}_3^{\text{NLO}}\left(x,\mu^2,P_z\right) 
	  & =\int_{|x|}^{\infty} d\xi\, 
	  \left[ C_3^{\text{NLO}}\left(\xi,\frac{\mu\xi}{xP_z}\right)
	  \frac{f_3\left(\frac{x}{\xi},\mu^2\right)}{|\xi|} - 
	  C_3^{\text{NLO}}\left(\xi,\frac{\mu}{xP_z}\right) f_3\left(x,\mu^2\right)\right] \nonumber \\
	  & \,\,\,\,\,\,\,\,\,\,\,\,\,\,\,\,\,\,\,\,\,\,\,\, 
	  + \int_{-\infty}^{-|x|}d\xi
	  \left[C_3^{\text{NLO}}\left(\xi,\frac{\mu\xi}{xP_z}\right)
	  \frac{f_3\left(\frac{x}{\xi},\mu^2\right)}{|\xi|} - 
	  C_3^{\text{NLO}}\left(\xi,\frac{\mu}{xP_z}\right)f_3\left(x,\mu^2\right)\right]. 
\end{align}
It can be easily verified that the two contributions appearing in the above
equation are indeed well defined for every fixed $x$: the singularity in
$\xi=+1$ is cured by the plus prescription, while for $\xi\rightarrow \pm
\infty$ the matching coefficient behaves like $C\left(\xi\right)\sim
\frac{1}{\xi^2}$, which is enough to guarantee the convergence of all the
integrals above. For numerical stability we find it useful to avoid the
singularity in $\xi = +1$ introducing a suitable small parameter $\delta \sim
10^{-6}$, and rewriting the above equation as
\begin{equation}
	\label{eq::explicitplus}
	\begin{split}
		\tilde{f}_3^{\text{NLO}}\left(x,\mu^2,P_z\right)
		&=\int_{|x|}^{1-\delta} d\xi\,
		C_3^{\text{NLO}}\left(\xi,\frac{\mu\xi}{xP_z}\right)
		\frac{f_3\left(\frac{x}{\xi},\mu\right)}{\xi} - 
		f_3\left(x,\mu\right) \int_{|x|}^{1-\delta} d\xi\,
		C_3^{\text{NLO}}\left(\xi,\frac{\mu}{xP_z}\right) \\
		& + \int_{1+\delta}^{\infty}d\xi\,
		C_3^{\text{NLO}}\left(\xi,\frac{\mu\xi}{xP_z}\right)
		\frac{f_3\left(\frac{x}{\xi},\mu\right)}{\xi} - 
		f_3\left(x,\mu\right) \int_{1+\delta}^{\infty}d\xi\,
		C_3^{\text{NLO}}\left(\xi,\frac{\mu}{xP_z}\right) \\
		& - \int_{-\infty}^{-|x|} d\xi\, 
		C_3^{\text{NLO}}\left(\xi,\frac{\mu\xi}{xP_z}\right)
		\frac{f_3\left(\frac{x}{\xi},\mu\right)}{\xi} - 
		f_3\left(x,\mu\right) \int_{-\infty}^{-|x|}d\xi\, 
		C_3^{\text{NLO}}\left(\xi,\frac{\mu}{xP_z}\right).
	\end{split}
\end{equation}
In order to obtain the lattice ME, we need to compute the real and imaginary part
of the Fourier transform of Eq.~\eqref{eq::explicitmatching} as shown in
Eq.~\eqref{eq:factorisedME}. Starting from the leading-order contribution, we get
\begin{align}
	\label{eq::LO}
	\int_{-\infty}^{\infty}& dx\, \cos\left(x P_z z\right) \int_{-1}^{1} \frac{dy}{|y|}\, \delta\lp 1-\frac{x}{y} \rp  
	f_{3}\lp y, {\mu}^2\rp  \nonumber \\
	&= \int_0^1 dy \cos\lp y P_z z  \rp \lp f_{3}\lp y, {\mu}^2\rp + f_{3}\lp -y, {\mu}^2\rp  \rp = 
	\int_0^1 dx \cos\lp x P_z z  \rp V_3 \lp x, {\mu}^2 \rp \nonumber \\
	&= \int_0^1 dx\, \text{A}^{\text{Re}, \,\text{LO}} \lp x P_z z \rp V_3 \lp x, {\mu}^2 \rp
\end{align}
where we have integrated in $x$ first, re-expressed the integral $\int_{-1}^1 dy$ as $\int_0^1 dy$, used
\begin{align} 
	f_3\left(x\right)+f_3\left(-x\right) = f_3^{\text{sym}}\left(x\right) = u^-\left(x\right) - d^-\left(x\right) = V_3\left(x\right) 
\end{align}
and finally changed variables back to $x$.  
Moving now to the next-to-leading order part, we analyse each of the six contributions listed in
Eq.~\eqref{eq::explicitplus}, defining for each lattice observable six integrals
to be computed, denoted as
$\text{I}^{\text{Re}}_i,\,\text{I}^{\text{Im}}_i\,\,\, i = 1,..,6$. Starting
from the first contribution to the real part we get
\begin{align}
	\text{I}^{\text{Re}}_1 
	  & =\int_{-\infty}^{\infty} dx\, \cos\left(x P_z z\right)
	  \int_{|x|}^{1-\delta}d\xi\, 
	  C_3^{\text{NLO}}\left(\xi,\frac{\mu\,\xi}{xP_z}\right)
	  \frac{f_3\left(\frac{x}{\xi},\mu\right)}{\xi} \nonumber \\
	  & = \int_{0}^{\infty} dx\, \cos\left(x P_z z\right)
	  \int_{x}^{1-\delta}\frac{d\xi}{\xi}\,
	  C_3^{\text{NLO}}\left(\xi,\frac{\mu\,\xi}{xP_z}\right)
	  \left(f_3\left(\frac{x}{\xi},\mu\right) + 
	  f_3\left(-\frac{x}{\xi},\mu\right) \right) \nonumber \\
	  & =\int_{0}^{1} dx\, \cos\left(x P_z z\right)
	  \int_{x/(1-\delta)}^{1}\frac{dy}{y}\,
	  C_3^{\text{NLO}}\left(\frac{x}{y},\frac{\mu}{y P_z}\right) V_3\left(y,\mu\right)\, ,
\end{align}
where in the last line we have changed variables back to $\smallfrac x \xi = y$.
Also, the integration range for $x$ becomes $\lp 0,1 \rp$, since $x<y<1$.
Renaming variables, we have
\begin{align}
	\text{I}^{\text{Re}}_1 
	  & = \int_{0}^{1}dx\,
	  	\left[
			  \frac{1}{x}\int_{0}^{1} dy\, 
			  \Theta\lp x - \smallfrac{y}{1-\delta} \rp
			  \cos\left(y P_z z\right) 
			  C_3^{\text{NLO}}\left(\frac{y}{x},\frac{\mu}{x P_z}\right)
		\right]
	V_3\left(x,\mu^2\right) \nonumber \\
	  & = \int_{0}^{1}dx\, 
		  \text{A}^{\text{Re},\,\text{NLO}}_1 \lp x, z, \frac{\mu}{P_z} \rp 
		  V_3\left(x,\mu^2\right)\, . 
\end{align}
Analogously, we find out that the other five contributions can be written as 
\begin{align}
	\text{I}^{\text{Re}}_{\,i} =
	\int_{0}^{1}dx\, \text{A}^{\text{Re},\,\text{NLO}}_{\,i} \lp x, z, \frac{\mu}{P_z} \rp V_3\left(x,\mu^2\right). 
\end{align}
with
\begin{align}
	& \text{A}^{\text{Re},\,\text{NLO}}_2 = \cos(x P_z z)\int_{x}^{1-\delta}d\xi \,
		C_3^{\text{NLO}}\left(\xi,\frac{\mu}{xP_z}\right), \\
	& \text{A}^{\text{Re},\,\text{NLO}}_3 = \frac{1}{x}\,\int_{0}^{\infty} dy\,
		\Theta\left(\smallfrac{y}{1+\delta}-x\right) \cos\left(y P_z z\right)C_3^{\text{NLO}}\left(\frac{y}{x},\frac{\mu}{x P_z}\right), \\
	& \text{A}^{\text{Re},\,\text{NLO}}_4 = \cos(x P_z z)\int_{1+\delta}^{\infty}d\xi\,
		C_3^{\text{NLO}}\left(\xi,\frac{\mu}{xP_z}\right), \\
	& \text{A}^{\text{Re},\,\text{NLO}}_5 = -\frac{1}{x}\,\int_{0}^{\infty} dy\,
		\cos\left(y P_z z\right) 
		C_3^{\text{NLO}}\left(-\frac{y}{x},\frac{\mu}{x P_z}\right), \\
	& \text{A}^{\text{Re},\,\text{NLO}}_6 = \cos(x P_z z) \int_{-\infty}^{-x}d\xi \,
		C_3^{\text{NLO}}\left(\xi,\frac{\mu}{xP_z}\right)
\end{align}
Collecting all the terms yields Eq.~\eqref{eq::V3factorization}
\begin{align}
	\mathcal{O}_{\gamma^0}^{\text{Re}}\lp z,\mu \rp = \int_{0}^{1} dx \, \mathcal{C}_3^{\text{Re}}\lp x, z, \frac{\mu}{P_z}\rp 
	\nsv\left(x,\mu^2\right)\, , 
\end{align}
where
\begin{align}
	\mathcal{C}_3^{\text{Re}}&\lp x, z, \frac{\mu}{P_z}  \rp = 
	\text{A}^{\text{Re},\,\text{LO}} + \text{A}^{\text{Re},\,\text{NLO}}  
\end{align}
with
\begin{align}
	\text{A}^{\text{Re},\,\text{NLO}}= 
	\text{A}^{\text{Re}, \,\text{NLO}}_1 - \text{A}^{\text{Re}, \,\text{NLO}}_2 + \text{A}^{\text{Re}, \,\text{NLO}}_3 
	- \text{A}^{\text{Re}, \,\text{NLO}}_4 - \text{A}^{\text{Re}, \,\text{NLO}}_5 - \text{A}^{\text{Re}, \,\text{NLO}}_6. 
\end{align}
We now turn to the imaginary part of the Fourier transform. The computation is
exactly the same as in the previous case, with the only difference that now we
have a $\sin$ instead of a $\cos$. Because of this, when re-expressing the
integral $\int_{-\infty}^{\infty} dx$ as $\int_{0	}^{\infty} dx $, we get an
additional minus sign, which gives the combination
\begin{align} 
	f\left(x\right)-f\left(-x\right) = f_3^{\text{asy}}\left(x\right) = u^+\left(x\right) - d^+\left(x\right) = T_3\left(x\right). 
\end{align} 
Therefore, the results for the imaginary part can be obtained from those for the real part simply by replacing $\cos$ with $\sin$ and $V_3$ with $T_3$.

\section{QCD evolution equations}
\label{app:dglap}
Let us briefly summarise how the QCD evolution equation is solved for the
nonsinglet sector, yielding the evolution kernel that is used in the
construction of FastKernel tables. For more details and for the validation of
this approach, we refer to Ref.~\cite{DelDebbio:2007ee} and the subsequent {\tt
NNPDF} publications.

Denoting the nonsiglet distributions $\nsv $ and $ \nst$ with $f^{(-)}$ and
$f^{(+)} $ respectively, the QCD evolution equations can be written as
\begin{align}
    \mu^2\frac{\partial }{\partial \mu^2} f^{(\pm)}\left(x,\mu^2\right) = 
    \frac{\alpha_s(\mu^2)}{2\pi}
    \int_{x}^{1}\frac{d\xi}{\xi}\, 
    P_{qq}^{(\pm)}\left(\frac{x}{\xi},\alpha_s(Q^2)\right)
    f^{(\pm)}\left(\xi,\mu^2\right),
\end{align}
which in Mellin space becomes
\begin{align}
\label{eq::dglapmellin}
    \mu^2\frac{\partial }{\partial \mu^2} f^{(\pm)}\left(N,\mu^2\right) = \gamma^{(\pm)}\left(N, \alpha_s\right) f^{(\pm)}\left(N,\mu^2\right)\, .
\end{align}
The distribution at the scale $\mu^2$ is obtained from the distribution at the
scale $\mu_0^2$ by introducing the evolution operator $\Gamma$
\begin{align}
\label{eq::evolutionoperator}
    f^{(\pm)}\left(N,\mu^2\right) = 
    \Gamma^{(\pm)}\left(N,\alpha_s,\alpha_s^0\right)
    f^{(\pm)}\left(N,\mu_0^2\right)\, ,
\end{align}
where $\alpha_s \equiv \alpha_s\left(\mu^2\right)$ and $\alpha_s^0 \equiv
\alpha_s\left(\mu_0^2\right)$. Substituting Eq.~\eqref{eq::evolutionoperator} in
Eq.~\eqref{eq::dglapmellin} and remembering that the dependence of $\Gamma$ on
the scale $\mu$ is only through the coupling, we have
\begin{align}
\label{eq::Mdglap}
    \beta\left(\alpha_s\right) \frac{\partial}{\partial\alpha_s}
    \Gamma^{(\pm)}\left(N,\alpha_s,\alpha_s^0\right) = 
    \gamma^{(\pm)}\left(N, \alpha_s\right)
    \Gamma^{(\pm)}\left(N,\alpha_s,\alpha_s^0\right)\, .
\end{align}
Since the matching coefficients in Eq.~\eqref{eq::matching} are known up to {\tt
NLO} ($\mathcal{O}\lp \alpha_s \rp$), here we will only consider {\tt NLO}
evolution equations. Then, using 
\begin{align}
    & \beta\left(\alpha_s\right) = \frac{d\alpha_s}{d\log \mu^2} = 
    -\alpha_s^2 \beta_0 -\alpha_s^3 \beta_1 + \mathcal{O}\lp \alpha_s^4 \rp  \\
    & \gamma^{(\pm)}\left(N, \alpha_s\right) = 
    \frac{\alpha_s}{4\pi} \gamma^{(\pm)}_0\left(N\right) + 
    \left(\frac{\alpha_s}{4\pi}\right)^2 \gamma^{(\pm)}_1\left(N\right) + \mathcal{O}\lp \alpha_s^3 \rp\, ,
\end{align}
we can solve Eq.~\eqref{eq::Mdglap}; the Mellin space expression for the evolution
kernel at {\tt NLO} is
\begin{align}
    \Gamma^{(\pm)}\left(N,\alpha_s,\alpha_s^0\right) = 
    1 + \frac{\alpha_s -\alpha_s^0}{4\pi}
    \left(\frac{\gamma^{(\pm)}_1 \left(N\right)}{2\beta_0} - 
    \frac{\beta_1 \gamma^{(\pm)}_0\left(N\right)}{2\beta_0^2}\right)
    \,.
\end{align}
The solution in the $x$-space is obtained by computing the inverse Mellin
transform of $\Gamma^{(\pm)}\left(N,\alpha_s,\alpha_s^0\right)$. Having
analytically continued the function $\Gamma\left(N\right)$ to the complex plane,
the inverse Mellin transform is obtained by computing the contour integral
\begin{align}
    \Gamma^{(\pm)}\left(x,\alpha_s,\alpha_s^0\right) = 
    \int_C \frac{dN}{2\pi i}x^{-N}\, 
    \Gamma^{(\pm)}\left(N,\alpha_s,\alpha_s^0\right)\, .
\end{align}


\bibliographystyle{UTPstyle}
\bibliography{main}

\end{document}